\documentclass[%
reprint,amsmath,amssymb,aps,prl,superscriptaddress,
twocolumn
]{revtex4-1}

\usepackage{graphicx}

\begin{document}
	
	\title{Mach-Zehnder interferometry using spin- and valley-polarized quantum Hall edge states in graphene}
	
	\author{Di S. Wei}
	\affiliation{John A. Paulson School of Engineering and Applied Sciences, Harvard University, Cambridge, Massachusetts 02138}
	
	\author{T. van der Sar}
	\affiliation{Department of Physics, Harvard University, Cambridge, Massachusetts 02138}
	
	\author{J. D. Sanchez-Yamagishi}
	\affiliation{Department of Physics, Harvard University, Cambridge, Massachusetts 02138}
	\affiliation{Department of Physics, Massachusetts Institute of Technology, Cambridge, Massachusetts 02139}
		
	\author{K. Watanabe}
	\affiliation{Advanced Materials Laboratory, National Institute for Materials Science, Tsukuba, Ibaraki 305-0044, Japan}
	
	\author{T. Taniguchi}
	\affiliation{Advanced Materials Laboratory, National Institute for Materials Science, Tsukuba, Ibaraki 305-0044, Japan}
	
	\author{P. Jarillo-Herrero}
	\affiliation{Department of Physics, Massachusetts Institute of Technology, Cambridge, Massachusetts 02139}
		
	\author{B.I. Halperin}
	\affiliation{Department of Physics, Harvard University, Cambridge, Massachusetts 02138}
		
	\author{A. Yacoby}
	\affiliation{Department of Physics, Harvard University, Cambridge, Massachusetts 02138}

	\maketitle
	
	\textbf{Confined to a two-dimensional plane, electrons in a strong magnetic field travel along the edge in one-dimensional quantum Hall channels that are protected against backscattering.  These channels can be used as solid-state analogues of monochromatic beams of light, providing a unique platform for studying electron interference. Electron interferometry is regarded as one of the most promising routes for studying fractional and non-Abelian statistics and quantum entanglement via two-particle interference. However, creating an edge-channel interferometer in which electron-electron interactions play an important role requires a clean system and long phase coherence lengths. Here we realize electronic Mach-Zehnder interferometers with record visibilities of up to 98$\%$ using spin- and valley-polarized edge channels that co-propagate along a PN junction in graphene. We find that inter-channel scattering between same-spin edge channels along the physical graphene edge can be used to form beamsplitters, while the absence of inter-channel scattering along gate-defined interfaces can be used to form isolated interferometer arms. Surprisingly, our interferometer is robust to dephasing effects at energies an order of magnitude larger than observed in pioneering experiments on GaAs/AlGaAs quantum wells. Our results shed light on the nature of edge-channel equilibration and open up new possibilities for studying exotic electron statistics and quantum phenomena.}  	
	\section{Introduction}
	Electron interference plays a central role in mesoscopic physics \cite{Datta1995, Chamon1997, Ji2003} and is regarded as one of the most promising routes for studying fractional and non-Abelian statistics \cite{Law2006, Feldman2006} and quantum entanglement via two-particle interference \cite{Samuelsson2004, Yurke1992}. Quantum Hall edges form excellent building blocks for electron interferometers as they are single-mode channels that are protected from inter-channel scattering by their quantum degrees of freedom such as spin \cite{Datta1995, Amet2014}. Furthermore, they can be positioned via electrostatic gating and coupled at target locations that act as beamsplitters \cite{Chamon1997,Ji2003}. Graphene may provide an advantage compared to conventional GaAs edge-channel interferometers \cite{Neder2006, Neder2007, Roulleau2007, Bieri2009} as the absence of a band gap allows the creation of hole- and electron-like edge channels that naturally meet, co-propagate, and separate at gate-defined PN interfaces \cite{Abanin2007, Williams2007}. Moreover, the additional valley degree of freedom and the associated unique nature of graphene quantum Hall states \cite{CastroNeto2009, Young2012} open up new opportunities for addressing long-sought goals of electron interferometry such as the observation of non-Abelian statistics \cite{Zibrov2017}. In addition, the valley isospin provides new possibilities for controlling inter-channel scattering \cite{Tworzydlo2007}, a requirement for creating edge-channel interferometers. However, even though graphene PN junctions in the quantum Hall regime have been studied extensively \cite{Amet2014, Abanin2007, Williams2007, Ozyilmaz2007, Velasco2009, Matsuo2015, Machida2015, Tovari2016}, creating an edge-channel interferometer using spin- and valley-polarized edge channels has remained an outstanding challenge. 
	
	In the paradigmatic electronic interferometer \textemdash{} the Mach-Zehnder interferometer (MZI) \cite{Ji2003} \textemdash{} a beam of electrons is split into two paths by a beam splitter and recombined at a second beam splitter.  Here, we engineer MZIs consisting of same-spin, opposite-valley quantum Hall edge channels that co-propagate along a PN junction in graphene. Using magnetic and trans-junction electric fields, we can tune into a regime in which either one or both pairs of the same-spin edge channels belonging to the zeroth Landau level (zLL) form MZIs that coherently mediate the cross-junction transport (Fig.~1A). We find that these channels can be well isolated from those belonging to other Landau levels (LLs), enabling us to study a target interferometer over a large range of electric fields and tune into regimes with visibilities as high as 98$\%$. By studying PN interfaces of different lengths, we show that the interferometer beamsplitters are located where the PN interface meets the physical graphene edges, which we attribute to strong inter-valley scattering at the physical graphene edge and the absence of inter-valley scattering along the gate-defined edge. We independently verify this conclusion using a device in which we can tune the number of edge channels co-propagating along either a physical or gate-defined edge.
	\section{Results}
	\subsection{Constructing a Mach-Zehnder interferometer in a graphene PN junction}
	
		\begin{figure}
			\includegraphics[width=0.5\textwidth]{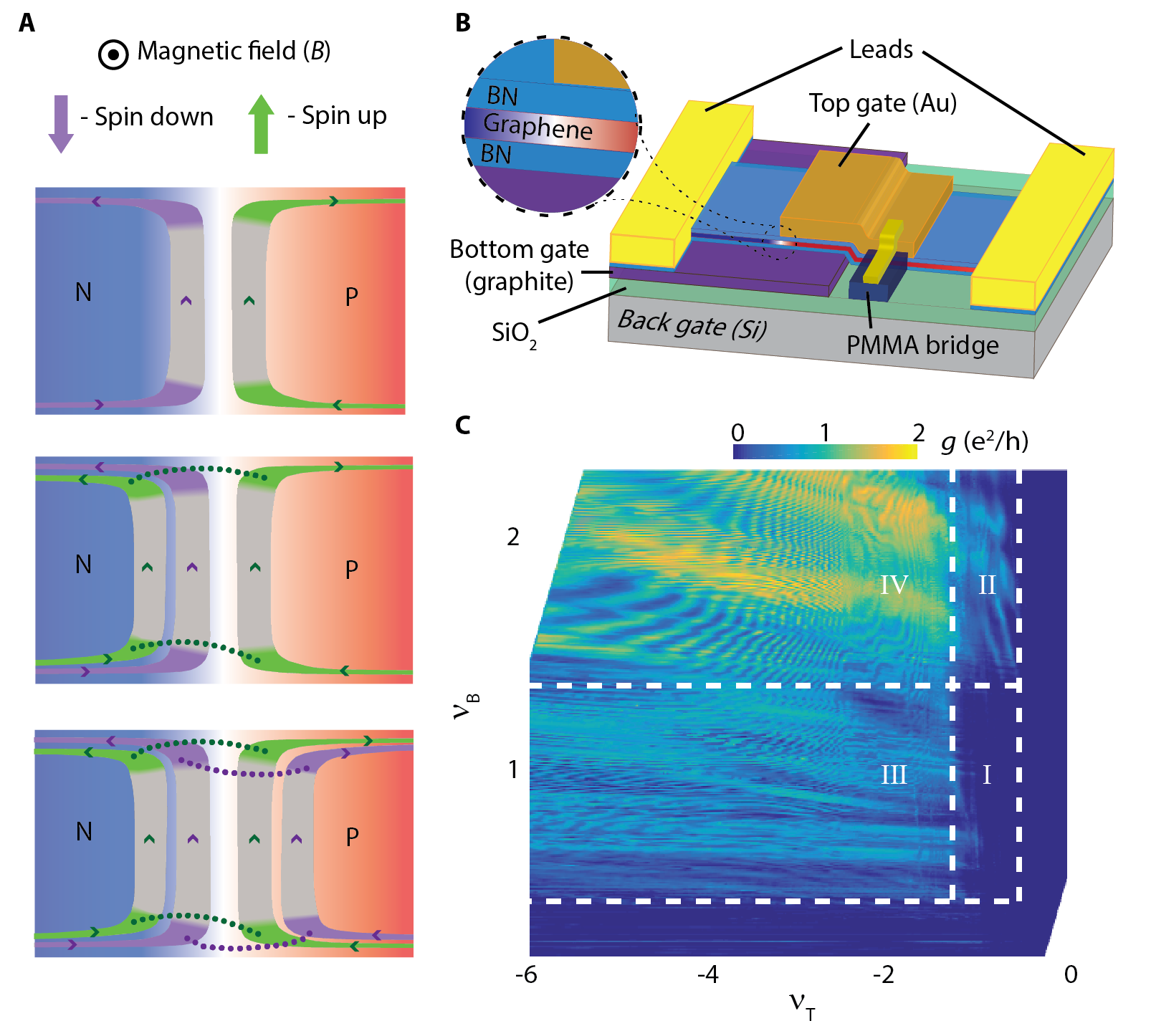}%
			\caption{\textbf{Creating a Mach-Zehnder interferometer using spin- and valley-polarized quantum Hall edge channels.} \textbf{(A)} Schematic illustration of the formation of Mach-Zehnder interferometers (MZIs) at a graphene PN junction. Green and purple denote quantum Hall edge channels of opposite spin. Top panel: at ($\nu_\mathrm{B},\nu_\mathrm{T}$)=(1,-1), where $\nu_\mathrm{B}$ ($\nu_\mathrm{T}$) is the filling factor in the N- (P-) region, two edge channels run along the interface. Their opposite spin suppresses inter-channel scattering. Middle panel: at ($\nu_\mathrm{B}$,$\nu_\mathrm{T}$)=(2,-1), a pair of spin-up edge channels forms a MZI. Inter-channel scattering occurs at the ends of the junction as indicated by dotted lines. Bottom panel: at ($\nu_\mathrm{B}$,$\nu_\mathrm{T}$)=(2,-2), two pairs of same-spin edge channels form two MZIs. \textbf{(B)} Device 1: an edge-contacted monolayer graphene flake encapsulated in hexagonal boron nitride (hBN). The top gate (Au) and bottom gate (graphite) define the PN junction (P: red color, N: blue color). The top (bottom) hBN gate dielectric is 20 (30) nm thick. The top gate is contacted by a lead that runs over a bridge fabricated from hard-baked PMMA to avoid shorting to the graphene flake. The back gate (Si) is used to strongly increase the p-doping of the graphene leading up to right lead and reduce the contact resistance. The $\mathrm{SiO}_2$ back-gate dielectric is 285 nm thick. \textbf{(C)} Two-terminal conductance of device 1 in the PN regime at \emph{B} = 4 T. We distinguish four regions (dashed boxes). Region I corresponds to ($\nu_\mathrm{B}$,$\nu_\mathrm{T}$) = (1,-1). Region II corresponds to $\nu_\mathrm{T} = -1$ and $\nu_\mathrm{B} \geq 2$. Region III corresponds to $\nu_\mathrm{B} = 1$ and $\nu_\mathrm{T} \leq -2$. Region IV corresponds to $\nu_\mathrm{B} \geq 2$ and $\nu_\mathrm{T} \leq -2$.}
			\label{fig:F1}
		\end{figure}
			
	To construct a MZI of spin- and valley-polarized edge channels, we use a hexagonal boron nitride (hBN) encapsulated monolayer of graphene (Fig.~1B and Methods). We tune into the quantum Hall regime using a perpendicular magnetic field \emph{B}, and define two regions of different charge densities $n_\mathrm{T}$ and $n_\mathrm{B}$ using a bottom gate that affects both $n_\mathrm{T}$ and $n_\mathrm{B}$ and a top gate that affects only $n_\mathrm{T}$ (Fig.~1B).  The number of edge channels in these regions is given by the filling factors $\nu_\mathrm{T,B}=(h/eB) n_\mathrm{T,B}$, where $e$ is the electron charge and $h$ is Planck's constant. The observation of integer quantum Hall steps in a measurement of the two-terminal conductance at \emph{B} = 4 T in the regime where $\nu_\mathrm{T}>0$ and $\nu_\mathrm{B}>0$ confirms that the spin- and valley-degeneracy is lifted (Supplementary Fig.~1). 
	
	Next, we create a PN junction by tuning into the regime where $\nu_\mathrm{T}<0$ and $\nu_\mathrm{B}>0$ and study which edge channels mediate charge transport across the junction. When we measure the conductance $g$ as a function of $\nu_\mathrm{T}<0$ and $\nu_\mathrm{B}>0$ at  \emph{B} = 4 T (Methods), we observe four regions with distinct ranges of conductance values, as well as the first indications of conductance oscillations (Fig.~1C). In region I, the conductance of the junction is near zero, which we attribute to the situation depicted in the top panel of Fig.~1A (where $\nu_\mathrm{B}=1$ and $\nu_\mathrm{T}=-1$). Here, one N-type spin-down and one P-type spin-up edge channel co-propagate along the junction. As these channels have opposite spin, inter-channel scattering is suppressed \cite{Amet2014}. When we cross from region I into region II, we begin to observe transport across the junction. We attribute this to an additional spin-up edge channel having entered on the N-side (so that $\nu_\mathrm{B}=2$ and $\nu_\mathrm{T}=-1$) and that electrons in this channel can scatter into the spin-up channel on the P-side (see middle panel in Fig.~1A). The observed conductance ranges approximately between 0 and $e^2/h$, consistent with one pair of edge channels mediating transport across the junction. Similarly, in region III we obtain the situation in which $\nu_\mathrm{B}=1$ and $\nu_\mathrm{T}=-2$, and we attribute the observed conductance to scattering between the two spin-down edge channels. Strikingly, in region III the conductance does not change notably as we keep adding edge channels on the P-side (going to $\nu_\mathrm{B}=1$ and $\nu_\mathrm{T}<-2$). We conclude that these additional channels do not contribute to the trans-junction conductance, presumably because they belong to a higher LL which makes them spatially too distant from the PN interface. Crossing into region IV ($\nu_\mathrm{B} \geq 2$ and  $\nu_\mathrm{T} \leq -2$), we observe that the average conductance increases and ranges between 0 and $2e^2/h$. We attribute this to two pairs of same-spin edge channels mediating transport across the junction. Again, we see no sign of edge channels belonging to higher LLs entering the system and contributing to the trans-junction conductance. We conclude that the edge channels belonging to the zLL mediate the trans-junction conductance, well isolated from edge channels belonging to higher LLs.
			
			\begin{figure}
				\includegraphics[width=0.5\textwidth]{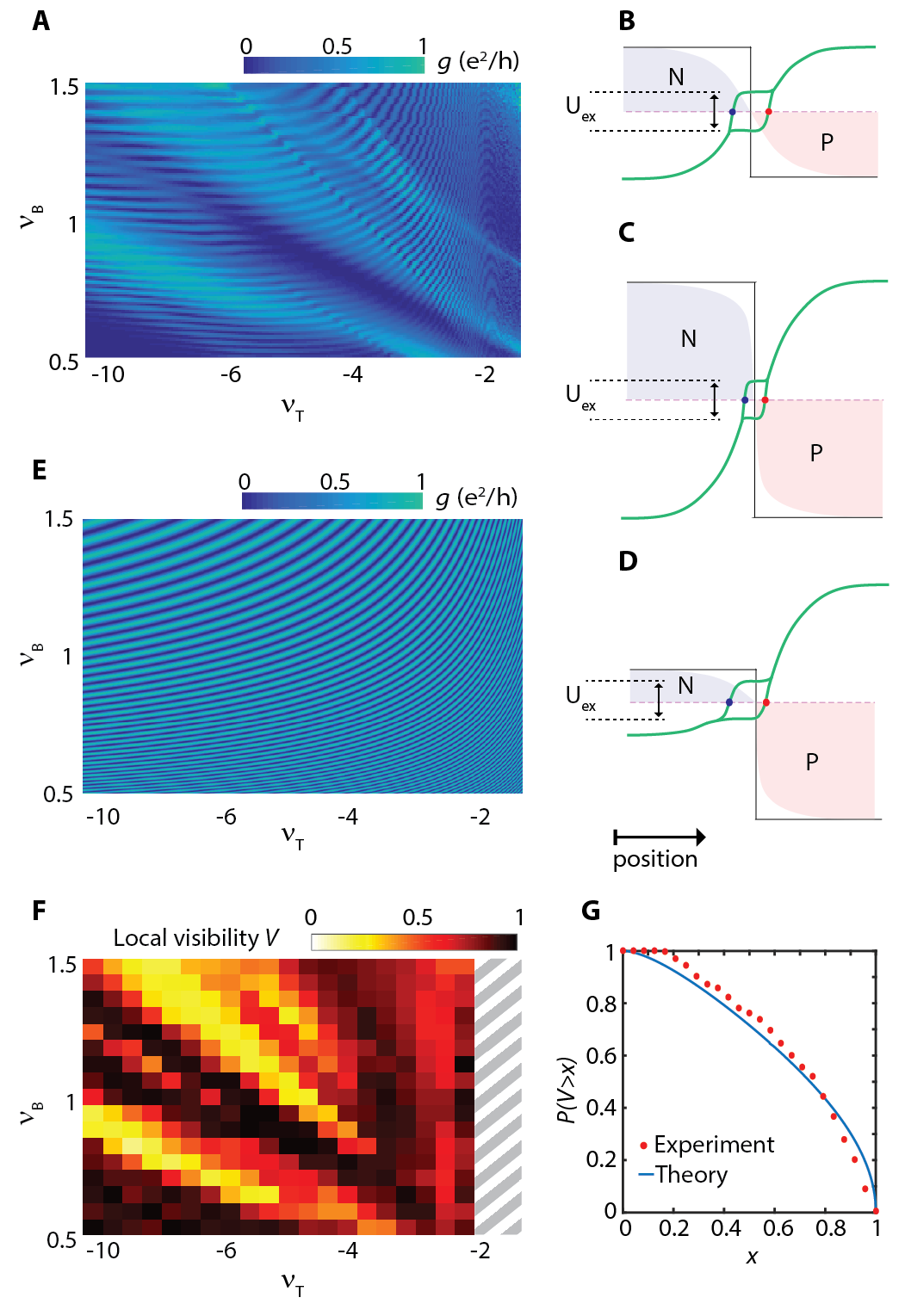}%
				\caption{\textbf{Characterization of a single Mach-Zehnder interferometer.} \textbf{(A)} Two-terminal conductance of device 1 at  \emph{B} = 9 T, over a range of filling factors corresponding to a single interferometer at the PN junction. \textbf{(B)} Modeling the charge-density dependence of the distance between the edge channels that form a MZI. The red and blue shading illustrates the spatial variation of the charge density close to the PN junction. The green line illustrates the spatial variation of the energies of the exchange-split $\nu$ = 1 and $\nu$ = -2 Landau sublevels. The edge channels are located at the positions where these sublevels intersect the Fermi energy. The distance between the edge channels determines the flux through the interferometer. Far from the PN junction, where the lowest LL is completely empty ($\nu$ = -2) or completely full ($\nu$ = 2), the exchange splitting $U_\mathrm{ex}$ vanishes. Near the PN junction, however, the electronic ground state can develop an imbalance in the valley occupation, leading self-consistently to a non-zero $U_\mathrm{ex}$. \textbf{(C)} Increasing the electron and hole densities decreases the distance between the edge channels. \textbf{(D)} A strong imbalance between the electron and hole densities. \textbf{(E)} Simulation of the two-terminal conductance as a function of filling factors based on the model sketched in (B-D). \textbf{(F)} Local visibility of the conductance oscillations observed in (A). The grey dashed box indicates where the visibility was not extracted due to non-resolved oscillations. \textbf{(G)} Data points: experimentally determined probability of finding a visibility greater than  \emph{x}, extracted from the color plot in (F). Blue line: theoretical prediction based on MZIs with beamsplitters described by random scattering matrices (Supplementary Note 2).}
				\label{fig:F2}
			\end{figure}
			
	The relative isolation of the edge channels that belong to the zLL allows us to study a target pair of edge channels over a large range of filling factors. As we increase the magnetic field to  \emph{B} = 9 T and concentrate on region III, in which $\nu_\mathrm{B}=1$ and $\nu_\mathrm{T} \leq -2$, we observe a striking pattern of conductance oscillations (Fig.~2A) whose key features such as shape and periodicity depend on both $\nu_\mathrm{B}$ and $\nu_\mathrm{T}$. These oscillations cannot be explained by semi-classical snake states or similar low-field phenomena \cite{Williams2011, Rickhaus2015, Taychatanapat2015, Overweg2016} since in our device electron transport is mediated by quantum Hall edge channels. Instead, as we will further argue below, the well-defined periodicity of these oscillations indicates that scattering between the two edge channels that mediate the cross-junction transport occurs at only two points along the junction. These points form the beamsplitters that define our MZI. Its conductance, in units of $e^2/h$, is given by  
	\begin{equation}
	g = |r_1t_2|^2+|t_1r_2|^2+2|t_1t_2r_1r_2| \text{cos}(\phi+\phi_0)
	\label{eq:cos}
	\end{equation}
	where $t_i$ ($r_i$) is the transmission (reflection) amplitude of the $i$-th beamsplitter, with $|r_i|^2+|t_i|^2=1$. The phase $\phi= \frac{2 \pi BA}{\Phi_0}$  arises from the Aharonov-Bohm effect, where $\Phi_0=h/e$ is the flux quantum, A the effective area enclosed by the two edge channels, and $\phi_0$ an (unknown) phase associated with the beamsplitters. 
	
	Since the measurement in Fig.~2A is performed at a fixed magnetic field, we attribute the conductance oscillations to a changing distance between the two edge channels and a resulting changing flux through the interferometer. We can analyze the charge-density-dependent locations of these channels by determining where the two corresponding exchange-split Landau sublevels cross the Fermi energy, using a simple model for the spatial dependence of the sublevel energy (Fig.~2B-D and Supplementary Note 1). This model indicates that as the charge densities increase (from Fig.~2B to Fig.~2C), the edge-channel separation decreases. Furthermore, when the charge density is small (large) on a particular side of the junction, the edge-channel separation is relatively sensitive (insensitive) to the charge density on that side of the junction (Fig.~2D). Figure 2E shows that this model reproduces the key features of the data in Fig.~2A. Further data in the $\nu_\mathrm{B} \geq 2$ and $\nu_\mathrm{T} \leq -2$ regime, in which two MZIs act simultaneously (as depicted in the bottom panel of Fig.~1A), are shown in Supplementary Fig.~2.
	\subsection{Beamsplitter characteristics}
	The visibility of the oscillations in a MZI depends on the phase coherence and the transmission characteristics of the beamsplitters.  We analyze the range of visibilities observed in the measurement shown in Fig.~2A by dividing the measurement range into a grid and calculating the local visibility $V=(g_\mathrm{max}-g_\mathrm{min})/(g_\mathrm{max}+g_\mathrm{min})$, with $g_\mathrm{max}$ and $g_\mathrm{min}$ the maximum and minimum conductance within each block (Fig.~2F). In Fig.~2G we plot the resulting experimental cumulative probability distribution function that indicates the probability of finding a visibility greater than $x$. This distribution corresponds well to a theoretical prediction that is based on the assumption that the incoming and outgoing channels of each of the two beamsplitters of the MZI are connected by random $U(2)$ matrices in valley space (Supplementary Note 2). Remarkably, in several regions of the conductance map (Fig.~2A) we find visibilities as high as 98$\%$, indicating near-perfect phase coherence along the PN interface. Additionally, in some regions the conductance oscillates nearly between 0 and $e^2/h$, indicating nearly 50/50 beam splitters.
	\subsection{Dependence of the Mach-Zehnder interference on magnetic field and DC voltage bias}
	Next, we tune to a region of high visibility and study the conductance as a function of \emph{B} and a DC voltage bias $V_\mathrm{DC}$ (Fig.~3A). We observe that the visibility stays near-unity for $|V_\mathrm{DC}|<0.5$ mV (Fig.~3B), and decreases at larger $|V_\mathrm{DC}|$, which may be due to thermal averaging or electron-electron interactions \cite{Ji2003, Bieri2009}. For the 8 to 9 T field range of Fig.~3C, measurements at $V_\mathrm{DC}=0$ show a constant oscillation period $\Delta B$, which is consistent with an assumption that the area enclosed by the interferometer is constant and given by $A=\frac{\Phi_0}{\Delta B}$. Subject to this assumption, we determine an edge-channel separation of 52 nm. Oscillations with $V_\mathrm{DC}$ are also observed (Fig.~3D) indicating a bias-dependent edge-channel separation, which may be a result of a bias-induced electrostatic gating effect \cite{Bieri2009}. We note that at larger filling factors we see multiple frequencies, changing frequency with field, and lobe structures, which have previously been attributed to Coulomb interactions in GaAs devices \cite{Neder2006, Roulleau2007,Bieri2009} (Supplementary Fig.~3). We leave the analysis of these effects to a future study.
		
			\twocolumngrid
						
			\begin{figure}
				\includegraphics[width=0.5\textwidth]{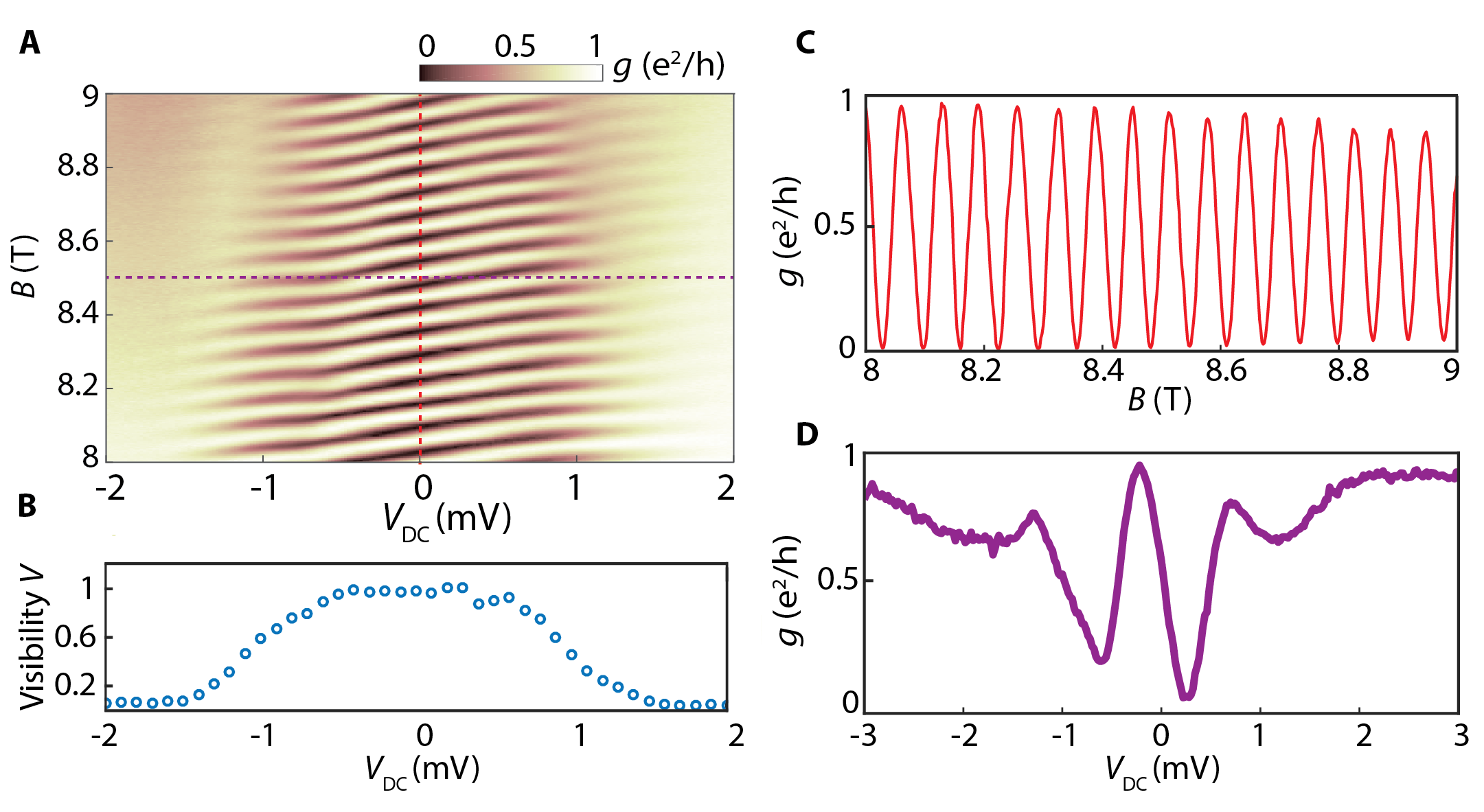}%
				\caption{\textbf{Mach-Zehnder oscillations as a function of magnetic field and DC voltage bias.} \textbf{(A)} Two-terminal differential conductance as a function of magnetic field \emph{B} and DC voltage bias $V_\mathrm{DC}$ at ($\nu_\mathrm{B}$,$\nu_\mathrm{T}$)=(1,-2), for which only one interferometer is formed at the PN interface. \textbf{(B)} Visibility of the conductance oscillations shown in (A) as a function of DC bias.\textbf{(C)} Conductance oscillations with \emph{B} at zero DC bias corresponding to the red dotted line in (A). From the period $\Delta B$ = 66 mT we calculate the distance between edge states to be 52 nm, assuming that the distance between the beamsplitters is given by the 1.2 $\mu$m width of the device. \textbf{(D)} Line trace corresponding to the purple dotted line in (A) showing oscillations with respect to $V_\mathrm{DC}$.}
							\label{fig:F3}
			\end{figure} 
			\twocolumngrid

	\twocolumngrid
	
		\begin{figure}[b] 
			\includegraphics[width=0.5\textwidth]{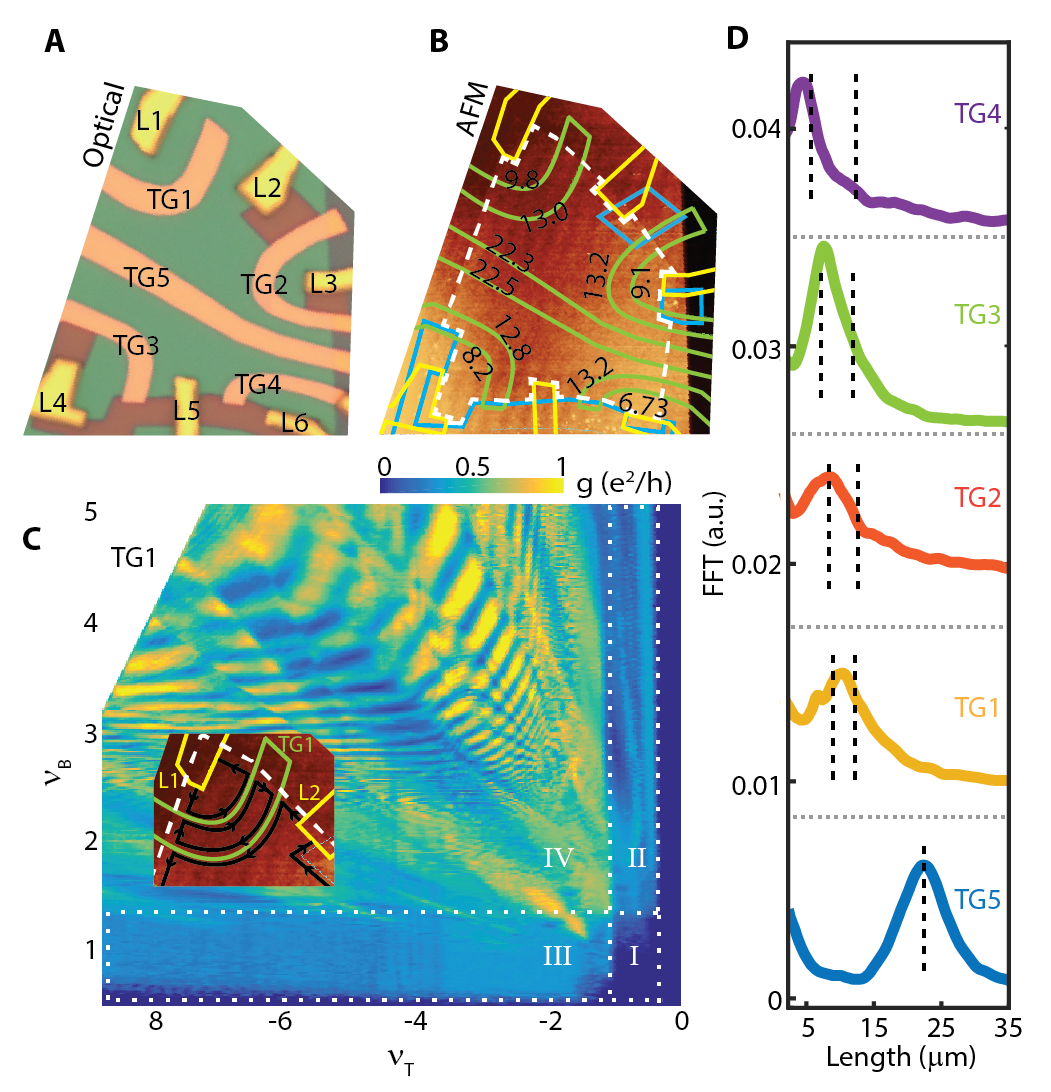}%
			\caption{\textbf{Gate-length dependence of the Mach-Zehnder oscillations.} \textbf{(A)} Optical microscope image of device 2: an edge-contacted, hBN-encapsulated monolayer of graphene with five top gates of different lengths. The top-gate dielectric (hBN) is 17 nm thick. The bottom hBN layer is 16 nm thick. The back-gate dielectric ($\mathrm{SiO}_2$) is 285 nm thick. Leads (L1-L6) are yellow. Top gates (TG1-TG5) are orange. Using  a top and back gate, we induce an NPN charge configuration with two PN junctions and their associated MZIs connected in series. \textbf{(B)} AFM image of device 2. The graphene is indicated by the dashed white line. Top gates are outlined in green, leads in yellow, etched regions in blue. The lengths of both sides of each top gate are indicated in micrometers. \textbf{(C)} Two-terminal conductance measured across top gate 1 (TG1) using leads L1 and L2 at \emph{B} = 8 T. Region I corresponds to ($\nu_\mathrm{B}$, $\nu_\mathrm{T}$) = (-1,1).  Region II corresponds to $\nu_\mathrm{T}$ = -1 and $\nu_\mathrm{B} \geq 2$. Region III corresponds to $\nu_\mathrm{T} \leq -2$ and $\nu_\mathrm{B}$ = 1. Region IV corresponds to $\nu_\mathrm{B} \geq 2$ and $\nu_\mathrm{T} \leq -2$. Inset: close-up of (B) showing the top gate and the two leads used in this measurement. The edge channels are indicated by black lines. \textbf{(D)} Frequency spectrum of the conductance oscillations for all top gates.  The x-axis is normalized to the length of TG5. The expected frequencies for each gate are indicated by the black dashed lines.}
			\label{fig:F4}
		\end{figure}
\twocolumngrid

	\begin{figure*}[t]
		\includegraphics[width=0.9\textwidth]{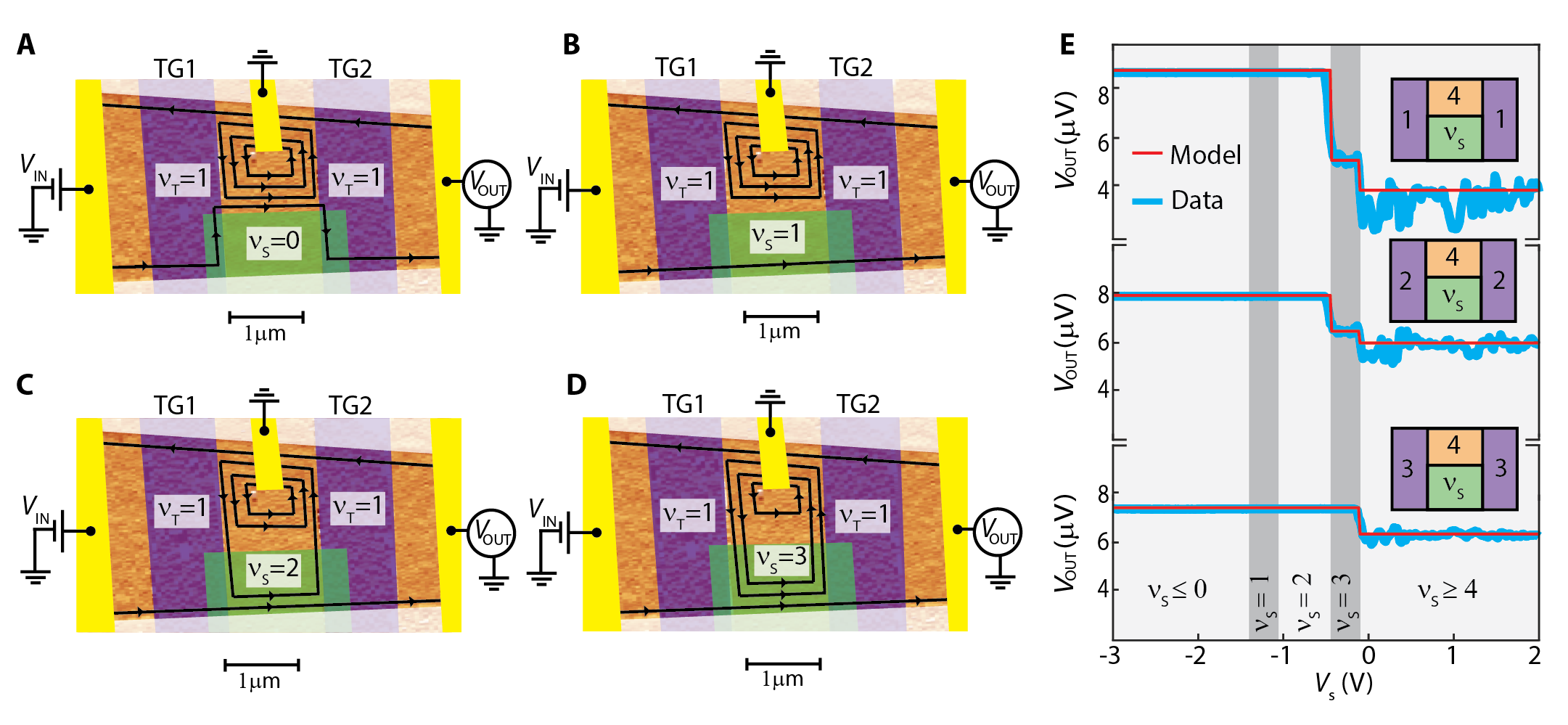}%
		\caption{\textbf{Absence of equilibration between edge channels running along a gate-defined edge.} \textbf{(A-D)} Schematic of device 3: an edge-contacted, hBN-encapsulated monolayer of graphene with two top gates and a side gate. An AFM image shows the top gates (TG1 and TG2, false-colored purple) on top of the hBN-encapsulated graphene flake. Yellow (green) indicates the leads (side gate). The progression of panels (A-D) illustrates the changing locations of the edge channels in the central region as the filling factor under the side gate is tuned from $\nu_\mathrm{S}$ = 0 to $\nu_\mathrm{S}$ = 3, while the regions under TG1 and TG2 are kept at $\nu_\mathrm{T}$ = 1 and the rest of the device is kept at $\nu_\mathrm{B}$ = 4. The circulating edge states near the contacts are omitted for clarity. \textbf{(E)} Voltage measured at the right contact as a function of the side-gate voltage $V_\mathrm{S}$ that tunes the side-gate filling factor $\nu_\mathrm{S}$, for $\nu_\mathrm{T}$ = 1, $\nu_\mathrm{T}$ = 2, and $\nu_\mathrm{T}$ = 3 as indicated by the insets. The data (blue) is an average of a set of traces taken at different magnetic fields between 7.9 and 8 T (Supplementary Fig.~6). The red line indicates the expected values given by a model that assumes no equilibration along the gate-defined edge and full equilibration between same-spin channels along the physical edge, taking into account the independently measured contact resistances (Supplementary Note 5). The top trace corresponds to the sequence depicted in (A-D).}
		\label{fig:F5}
	\end{figure*}
\subsection{Varying the length of the PN interface}
	To confirm that the beamsplitters are located where the PN interface meets the physical graphene edges, we measure the MZI oscillation frequency as a function of the interface length. We use device 2 (Fig.~4A-B), which has five top gates (TG1 to TG5) of varying lengths that we can address individually in two-terminal conductance measurements by using the appropriate leads. Using top and bottom gates to control the filling factors in the top-gated and non-top-gated regions, $\nu_\mathrm{T}$ and $\nu_\mathrm{B}$ respectively, we can tune into a regime where $\nu_\mathrm{T}<0$ and $\nu_\mathrm{B}>0$ to create an NPN configuration with two PN junctions in series (Supplementary Fig.~4). When we measure the two-terminal conductance at \emph{B} = 8 T as a function of  $\nu_\mathrm{T}$ and $\nu_\mathrm{B}$ (Fig.~4C and Supplementary Fig.~4), we recognize the regions corresponding to zero, one, and two pairs of same-spin edge channels mediating transport across the PN junctions, as discussed above for the measurement in Fig.~1C and further analyzed in Supplementary Note 3. In addition, we observe clear conductance oscillations, of which we expect the frequencies to reflect the gate lengths. To analyze these frequencies, we focus on the limit $|\nu_\mathrm{B,T}| >> 1$ in which the edge-channel separation and the associated Aharanov-Bohm flux are expected to vary as $ \sim 1/(\sqrt{\nu_\mathrm{B}}+\sqrt{-\nu_\mathrm{T}}$) (Supplementary Note 4). We plot the conductance data against $1/(\sqrt{\nu_\mathrm{B}}+\sqrt{-\nu_\mathrm{T}}$) and use a Fourier transform to determine the frequency spectrum (Supplementary Fig.~5). Normalizing the frequency axis to the average length of TG5, we find peaks at locations that correspond reasonably well to those expected based on the lengths of the different gates (Fig.~4D). We conclude that the beamsplitters are located where the PN interfaces meet the physical graphene edge. Remarkably, it follows that each oscillation corresponds to a minute change in the edge-channel separation: for example, for the L=1.2 $\mu$m gate length of device 1, this change equals $\frac{\Phi_0}{BL}=3.7$ \AA.
	\subsection{Edge-channel equilibration along gate-defined and physical edges}
	Finally, we demonstrate the absence of inter-channel scattering along a gate-defined edge and the full equilibration of same-spin edge channels running along a physical edge. We use device 3 (Fig.~5A-D), which has two top gates that determine the number of edge channels running from the left to the right lead, and a top gate (referred to as the \emph{side gate}) that determines which fraction of the edge channels in the central region travel along the lower physical edge instead of along the side-gate-defined edge. We first confirm the presence of robust broken-symmetry quantum Hall states (Supplementary Fig.~6C). We then apply a bias $V_\mathrm{IN}$ between the left and top lead, and measure the potential at the right lead ($V_\mathrm{OUT}$) as a function of the side-gate filling factor (Fig.~5E). Edge-channel equilibration in the central region should reduce the chemical potential at the right lead below that of the input lead. The precise match between data and model (described in Supplementary Note 5) in Fig.~5E clearly demonstrates that edge channels do not equilibrate along the side-gate-defined edge, while they do equilibrate along the physical edge provided they have the same spin \cite{Amet2014}. We note that we observe no MZI oscillations as we sweep the magnetic field between 8.9 and 9 T (Supplementary Fig.~6D-F), presumably because there are no locations acting as beamsplitters as the edge channels do not meet at a physical edge before and after co-propagating along a gate-defined edge.
	\section{Discussion}
	The experiments presented here demonstrate a robust method of engineering a high-visibility MZI in a graphene QH system. This opens up the possibility of a variety of interferometry experiments and grants us the diagnostic capabilities of measuring sub-nanometer shifts in edge-channel separation. In our experiments we observe transport across the insulating $\nu=0$ state, which is expected to be in a canted antiferromagnetic (CAF) phase in bulk graphene \cite{Kharitonov2012, Young2014}. The fact that we find that spin polarization is well preserved in our samples suggests that the CAF phase may be suppressed in a narrow PN junction in favor of a state where spins are fully polarized along the direction of the magnetic field. 
	\section{Methods} 
	\subsection{Sample Fabrication} All devices were fabricated on doped Si chips with a 285 nm layer of $\mathrm{SiO}_2$ that acted as a dielectric for the Si back gate. Graphene was mechanically exfoliated from bulk graphite obtained from NGS Naturgraphit GmbH using 1009R tape from Ultron Systems and subsequently encapsulated in hexagonal boron nitride (hBN) using a dry transfer process \cite{Wang2013}. For device 1, we placed the resulting stack on a graphite bottom gate. Before the first metal deposition step, we annealed the devices in vacuum at 500$^\circ$C to improve device quality. We then created top gates using electron-beam lithography and thermal evaporation of Cr/Au. To fabricate edge-contacts to the graphene in device 1 without shorting to the graphite bottom gate, we selectively etched the stack down such that the bottom hBN flake remained and protected the graphite while simultaneously exposing the graphene flake. To fabricate edge-contacts to the graphene in devices 2 and 3, we etched through the entire hBN/graphene stack. We then created edge contacts by thermally evaporating Cr/Au while rotating the sample using a tilted rotation stage. Finally, we etched the devices into the desired geometry by reactive ion etching in $\mathrm{O}_2/\mathrm{CHF}_3$ using a PMMA/HSQ bilayer of resist (patterned by electron-beam lithography) as the etch mask.
	\subsection{Measurement} Our measurements were performed in a Leiden dry dilution refrigerator with a base temperature of 20 mK. Measurements of differential conductance were performed using a lock-in amplifier with an AC excitation voltage of 10 $\mu$V at 17.77 Hz. All measurements of differential conductance were corrected for contact/line resistances, which were independently determined by lining up the robust $\nu = 2$ quantum Hall conductance plateau with $2e^2/h$. We estimated all filling factors based on a parallel-plate capacitor model with a correction to account for quantum capacitance (Supplementary Note 6).
	\bibliographystyle{naturemag}
	\bibliography{Bibliography}

\begin{thebibliography}{10}
\expandafter\ifx\csname url\endcsname\relax
  \def\url#1{\texttt{#1}}\fi
\expandafter\ifx\csname urlprefix\endcsname\relax\def\urlprefix{URL }\fi
\providecommand{\bibinfo}[2]{#2}
\providecommand{\eprint}[2][]{\url{#2}}

\bibitem{Datta1995}
\bibinfo{author}{Datta, S.}
\newblock \emph{\bibinfo{title}{Electronic Transport in Mesoscopic Systems}}
  (\bibinfo{publisher}{Cambridge Univ. Press}, \bibinfo{year}{1995}).

\bibitem{Chamon1997}
\bibinfo{author}{Chamon, C.~C.}, \bibinfo{author}{Freed, D.~E.},
  \bibinfo{author}{Kivelson, S.~A.}, \bibinfo{author}{Sondhi, S.~L.} \&
  \bibinfo{author}{Wen, X.~G.}
\newblock \bibinfo{title}{Two point-contact interferometer for quantum {Hall}
  systems}.
\newblock \emph{\bibinfo{journal}{Phys. Rev. B}} \textbf{\bibinfo{volume}{55}},
  \bibinfo{pages}{2331--2343} (\bibinfo{year}{1997}).

\bibitem{Ji2003}
\bibinfo{author}{Ji, Y.}, \bibinfo{author}{Chung, Y.},
  \bibinfo{author}{Sprinzak, D.}, \bibinfo{author}{Heiblum, M.} \&
  \bibinfo{author}{Mahalu, D.}
\newblock \bibinfo{title}{{An electronic Mach{-}Zehnder interferometer}}.
\newblock \emph{\bibinfo{journal}{Nature}} \textbf{\bibinfo{volume}{422}},
  \bibinfo{pages}{415--418} (\bibinfo{year}{2003}).

\bibitem{Law2006}
\bibinfo{author}{Law, K.~T.}, \bibinfo{author}{Feldman, D.~E.} \&
  \bibinfo{author}{Gefen, Y.}
\newblock \bibinfo{title}{Electronic {M}ach-{Z}ehnder interferometer as a tool
  to probe fractional statistics}.
\newblock \emph{\bibinfo{journal}{Phys. Rev. B}} \textbf{\bibinfo{volume}{74}},
  \bibinfo{pages}{045319} (\bibinfo{year}{2006}).

\bibitem{Feldman2006}
\bibinfo{author}{Feldman, D.~E.} \& \bibinfo{author}{Kitaev, A.}
\newblock \bibinfo{title}{Detecting non-abelian statistics with an electronic
  {M}ach-{Z}ehnder interferometer}.
\newblock \emph{\bibinfo{journal}{Phys. Rev. Lett.}}
  \textbf{\bibinfo{volume}{97}}, \bibinfo{pages}{186803}
  (\bibinfo{year}{2006}).

\bibitem{Samuelsson2004}
\bibinfo{author}{Samuelsson, P.}, \bibinfo{author}{Sukhorukov, E.~V.} \&
  \bibinfo{author}{B\"uttiker, M.}
\newblock \bibinfo{title}{Two-particle {A}haronov-{B}ohm effect and
  entanglement in the electronic {H}anbury {B}rown-{T}wiss setup}.
\newblock \emph{\bibinfo{journal}{Phys. Rev. Lett.}}
  \textbf{\bibinfo{volume}{92}}, \bibinfo{pages}{026805}
  (\bibinfo{year}{2004}).

\bibitem{Yurke1992}
\bibinfo{author}{Yurke, B.} \& \bibinfo{author}{Stoler, D.}
\newblock \bibinfo{title}{Bell's-inequality experiments using
  independent-particle sources}.
\newblock \emph{\bibinfo{journal}{Phys. Rev. A}} \textbf{\bibinfo{volume}{46}},
  \bibinfo{pages}{2229--2234} (\bibinfo{year}{1992}).

\bibitem{Amet2014}
\bibinfo{author}{Amet, F.}, \bibinfo{author}{Williams, J.~R.},
  \bibinfo{author}{Watanabe, K.}, \bibinfo{author}{Taniguchi, T.} \&
  \bibinfo{author}{Goldhaber-Gordon, D.}
\newblock \bibinfo{title}{Selective equilibration of spin-polarized quantum
  {H}all edge states in graphene}.
\newblock \emph{\bibinfo{journal}{Phys. Rev. Lett.}}
  \textbf{\bibinfo{volume}{112}}, \bibinfo{pages}{196601}
  (\bibinfo{year}{2014}).

\bibitem{Neder2006}
\bibinfo{author}{Neder, I.}, \bibinfo{author}{Heiblum, M.},
  \bibinfo{author}{Levinson, Y.}, \bibinfo{author}{Mahalu, D.} \&
  \bibinfo{author}{Umansky, V.}
\newblock \bibinfo{title}{{Unexpected behavior in a two-path electron
  interferometer}}.
\newblock \emph{\bibinfo{journal}{Physical Review Letters}}
  \textbf{\bibinfo{volume}{96}}, \bibinfo{pages}{1--4} (\bibinfo{year}{2006}).

\bibitem{Neder2007}
\bibinfo{author}{Neder, I.} \emph{et~al.}
\newblock \bibinfo{title}{{Interference between two indistinguishable electrons
  from independent sources.}}
\newblock \emph{\bibinfo{journal}{Nature}} \textbf{\bibinfo{volume}{448}},
  \bibinfo{pages}{333--337} (\bibinfo{year}{2007}).

\bibitem{Roulleau2007}
\bibinfo{author}{Roulleau, P.} \emph{et~al.}
\newblock \bibinfo{title}{{Finite bias visibility of the electronic
  {M}ach-{Z}ehnder interferometer}}.
\newblock \emph{\bibinfo{journal}{Physical Review B}}
  \textbf{\bibinfo{volume}{76}}, \bibinfo{pages}{161309}
  (\bibinfo{year}{2007}).

\bibitem{Bieri2009}
\bibinfo{author}{Bieri, E.} \emph{et~al.}
\newblock \bibinfo{title}{Finite-bias visibility dependence in an electronic
  {M}ach-{Z}ehnder interferometer}.
\newblock \emph{\bibinfo{journal}{Phys. Rev. B}} \textbf{\bibinfo{volume}{79}},
  \bibinfo{pages}{245324} (\bibinfo{year}{2009}).

\bibitem{Abanin2007}
\bibinfo{author}{Abanin, D.~A.} \& \bibinfo{author}{Levitov, L.~S.}
\newblock \bibinfo{title}{{Quantized transport in graphene p-n junctions in a
  magnetic field.}}
\newblock \emph{\bibinfo{journal}{Science}} \textbf{\bibinfo{volume}{317}},
  \bibinfo{pages}{641--643} (\bibinfo{year}{2007}).

\bibitem{Williams2007}
\bibinfo{author}{Williams, J.~R.}, \bibinfo{author}{Dicarlo, L.} \&
  \bibinfo{author}{Marcus, C.~M.}
\newblock \bibinfo{title}{{Quantum Hall effect in a gate-controlled p-n
  junction of graphene}}.
\newblock \emph{\bibinfo{journal}{Science}} \textbf{\bibinfo{volume}{317}},
  \bibinfo{pages}{638--640} (\bibinfo{year}{2007}).

\bibitem{CastroNeto2009}
\bibinfo{author}{Castro~Neto, A.~H.}, \bibinfo{author}{Guinea, F.},
  \bibinfo{author}{Peres, N. M.~R.}, \bibinfo{author}{Novoselov, K.~S.} \&
  \bibinfo{author}{Geim, A.~K.}
\newblock \bibinfo{title}{The electronic properties of graphene}.
\newblock \emph{\bibinfo{journal}{Rev. Mod. Phys.}}
  \textbf{\bibinfo{volume}{81}}, \bibinfo{pages}{109--162}
  (\bibinfo{year}{2009}).

\bibitem{Young2012}
\bibinfo{author}{Young, A.~F.} \emph{et~al.}
\newblock \bibinfo{title}{{Spin and valley quantum Hall ferromagnetism in 
  graphene}}.
\newblock \emph{\bibinfo{journal}{Nature Physics}}
  \textbf{\bibinfo{volume}{8}}, \bibinfo{pages}{550--556}
  (\bibinfo{year}{2012}).

\bibitem{Zibrov2017}
\bibinfo{author}{Zibrov, A.} \emph{et~al.}
\newblock \bibinfo{title}{Robust nonabelian ground states and continuous
  quantum phase transitions in a half filled bilayer graphene {L}andau level}.
\newblock \emph{\bibinfo{journal}{Preprint at
  https://arxiv.org/abs/1611.07113}}  (\bibinfo{year}{2016}).

\bibitem{Tworzydlo2007}
\bibinfo{author}{Tworzyd\l{}o, J.}, \bibinfo{author}{Snyman, I.},
  \bibinfo{author}{Akhmerov, A.~R.} \& \bibinfo{author}{Beenakker, C. W.~J.}
\newblock \bibinfo{title}{Valley-isospin dependence of the quantum {H}all
  effect in a graphene $p\text{-}n$ junction}.
\newblock \emph{\bibinfo{journal}{Phys. Rev. B}} \textbf{\bibinfo{volume}{76}},
  \bibinfo{pages}{035411} (\bibinfo{year}{2007}).

\bibitem{Ozyilmaz2007}
\bibinfo{author}{{\"{O}}zyilmaz, B.} \emph{et~al.}
\newblock \bibinfo{title}{{Electronic transport and quantum Hall effect in
  bipolar graphene p-n-p Junctions}}.
\newblock \emph{\bibinfo{journal}{Phys. Rev. Lett.}}
  \textbf{\bibinfo{volume}{99}}, \bibinfo{pages}{166804}
  (\bibinfo{year}{2007}).

\bibitem{Velasco2009}
\bibinfo{author}{Velasco, J.}, \bibinfo{author}{Liu, G.}, \bibinfo{author}{Bao,
  W.} \& \bibinfo{author}{{Ning Lau}, C.}
\newblock \bibinfo{title}{{Electrical transport in high-quality graphene pnp
  junctions}}.
\newblock \emph{\bibinfo{journal}{New Journal of Physics}}
  \textbf{\bibinfo{volume}{11}}, \bibinfo{pages}{095008}
  (\bibinfo{year}{2009}).

\bibitem{Matsuo2015}
\bibinfo{author}{Matsuo, S.} \emph{et~al.}
\newblock \bibinfo{title}{{Edge mixing dynamics in graphene p-n junctions in
  the quantum Hall regime}}.
\newblock \emph{\bibinfo{journal}{Nature Communications}}
  \textbf{\bibinfo{volume}{6}}, \bibinfo{pages}{8066} (\bibinfo{year}{2015}).

\bibitem{Machida2015}
\bibinfo{author}{Machida, T.} \emph{et~al.}
\newblock \bibinfo{title}{{Edge-channel transport of Dirac fermions in graphene
  quantum {H}all junctions}}.
\newblock \emph{\bibinfo{journal}{Journal of the Physical Society of Japan}}
  \textbf{\bibinfo{volume}{84}}, \bibinfo{pages}{1--9} (\bibinfo{year}{2015}).

\bibitem{Tovari2016}
\bibinfo{author}{Tovari, E.} \emph{et~al.}
\newblock \bibinfo{title}{Gate-controlled conductance enhancement from quantum
  {H}all channels along graphene p-n junctions}.
\newblock \emph{\bibinfo{journal}{Nanoscale}} \textbf{\bibinfo{volume}{8}},
  \bibinfo{pages}{19910--19916} (\bibinfo{year}{2016}).

\bibitem{Williams2011}
\bibinfo{author}{Williams, J.~R.} \& \bibinfo{author}{Marcus, C.~M.}
\newblock \bibinfo{title}{Snake states along graphene $p\mathrm{\text{-}}n$
  junctions}.
\newblock \emph{\bibinfo{journal}{Phys. Rev. Lett.}}
  \textbf{\bibinfo{volume}{107}}, \bibinfo{pages}{046602}
  (\bibinfo{year}{2011}).

\bibitem{Rickhaus2015}
\bibinfo{author}{Rickhaus, P.} \emph{et~al.}
\newblock \bibinfo{title}{{Snake trajectories in ultraclean graphene p–n
  junctions}}.
\newblock \emph{\bibinfo{journal}{Nature Communications}}
  \textbf{\bibinfo{volume}{6}}, \bibinfo{pages}{6470} (\bibinfo{year}{2015}).

\bibitem{Taychatanapat2015}
\bibinfo{author}{Taychatanapat, T.} \emph{et~al.}
\newblock \bibinfo{title}{{Conductance oscillations induced by ballistic snake
  states in a graphene heterojunction}}.
\newblock \emph{\bibinfo{journal}{Nature Communications}}
  \textbf{\bibinfo{volume}{6}}, \bibinfo{pages}{6093} (\bibinfo{year}{2015}).

\bibitem{Overweg2016}
\bibinfo{author}{Overweg, H.} \emph{et~al.}
\newblock \bibinfo{title}{{Oscillating magnetoresistance in graphene p-n
  junctions at intermediate magnetic fields}}.
\newblock \emph{\bibinfo{journal}{Preprint at
  https://arxiv.org/abs/1612.07624}}  (\bibinfo{year}{2016}).

\bibitem{Kharitonov2012}
\bibinfo{author}{Kharitonov, M.}
\newblock \bibinfo{title}{Phase diagram for the $\ensuremath{\nu}=0$ quantum
  hall state in monolayer graphene}.
\newblock \emph{\bibinfo{journal}{Phys. Rev. B}} \textbf{\bibinfo{volume}{85}},
  \bibinfo{pages}{155439} (\bibinfo{year}{2012}).

\bibitem{Young2014}
\bibinfo{author}{Young, A.~F.} \emph{et~al.}
\newblock \bibinfo{title}{{Tunable symmetry breaking and helical edge transport
  in a graphene quantum spin Hall state.}}
\newblock \emph{\bibinfo{journal}{Nature}} \textbf{\bibinfo{volume}{505}},
  \bibinfo{pages}{528--32} (\bibinfo{year}{2014}).

\bibitem{Wang2013}
\bibinfo{author}{Wang, L.} \emph{et~al.}
\newblock \bibinfo{title}{{One-dimensional electrical contact to a
  two-dimensional material.}}
\newblock \emph{\bibinfo{journal}{Science}} \textbf{\bibinfo{volume}{342}},
  \bibinfo{pages}{614--7} (\bibinfo{year}{2013}).

\bibitem{Zhang2005}
\bibinfo{author}{Zhang, Y.}, \bibinfo{author}{Tan, Y.-W.},
  \bibinfo{author}{Stormer, H.~L.} \& \bibinfo{author}{Kim, P.}
\newblock \bibinfo{title}{{Experimental observation of the quantum Hall effect
  and Berry's phase in graphene.}}
\newblock \emph{\bibinfo{journal}{Nature}} \textbf{\bibinfo{volume}{438}},
  \bibinfo{pages}{201--204} (\bibinfo{year}{2005}).

\bibitem{McClure2009}
\bibinfo{author}{McClure, D.~T.} \emph{et~al.}
\newblock \bibinfo{title}{{Edge-state velocity and coherence in a quantum Hall
  Fabry-P\'erot interferometer}}.
\newblock \emph{\bibinfo{journal}{Phys. Rev. Lett.}}
  \textbf{\bibinfo{volume}{103}}, \bibinfo{pages}{206806}
  (\bibinfo{year}{2009}).

\end{thebibliography}
	\section{Acknowledgements}
	\subsection{General}
	We acknowledge helpful discussions and feedback from  A. Kou, B.E. Feldman, D. Nandi, K. Shain, P. Kim, A. Stern, M. Heiblum, C. Schonenberger, K. Wang, S.L. Tomarken and S.P. Harvey. We also thank G.H. Lee, J.I.J. Wang, J.Y. Luo and X. Liu for help with fabrication.
	\subsection{Funding}This work was primarily supported by the US DOE, BES Office, Division of Materials Sciences and Engineering under Award DE-SC0001819 (D.S.W., J.D.S.Y., P.J.H. and A.Y.). D.S.W. acknowledges partial support from the National Science Foundation Graduate Research Fellowship under Grant No. DGE1144152. B.I.H. acknowledges support from the STC Center for Integrated Quantum Materials, NSF grant DMR-1231319. K.W. and T.T. acknowledge support from the Elemental Strategy Initiative conducted by the MEXT, Japan and JSPS KAKENHI Grant Numbers JP26248061, JP15K21722 and JP25106006. This work was performed in part at the Harvard University Center for Nanoscale Systems (CNS), a member of the National Nanotechnology Coordinated Infrastructure Network (NNCI), which is supported by the National Science Foundation under NSF ECCS award no. 1541959. 
	\subsection{Author Contributions} D.S.W., T.S., B.I.H. and A.Y. conceived and designed the experiments. D.S.W. fabricated the devices. D.S.W., T.S., J.D.S.Y. and A.Y. performed the experiments. D.S.W., T.S., J.D.S.Y., P.J.-H., B.I.H. and A.Y. analyzed the data, developed the models, and wrote the paper. K.W. and T.T. provided the hexagonal boron nitride crystals used in the devices. 
	\section{Correspondence} Correspondence and requests for materials should be addressed to D.S.W. (email: \mbox{\texttt{diwei@fas.harvard.edu}}).	
	\newcommand{\beginsupplement}{%
		\setcounter{table}{0}
		\renewcommand{\thetable}{S\arabic{table}}%
		\setcounter{figure}{0}
		\renewcommand{\thefigure}{S\arabic{figure}}%
		\setcounter{equation}{0}
		\renewcommand{\theequation}{S\arabic{equation}}
	}
		\beginsupplement
		
		\section{Supplementary Information}
		
		
		\subsection{Supplementary Note 1. Modeling the distance between the edge channels forming a Mach-Zehnder interferometer}
		
		In this note we describe the model used for the calculations shown in Fig.~2E of the main text, which shows the conductance of a PN junction as a function of filling factors to the left and right of the PN junction, $\nu_\mathrm{B}$ and $\nu_\mathrm{T}$, respectively.  As discussed in the main text, we attribute the oscillations to a changing flux enclosed by the two edge channels forming a Mach-Zehnder interferometer. At fixed magnetic field, this change is caused by a change in distance between the edge channels. We analyze the locations of the two interferometer-forming edge channels by determining where the corresponding exchange-split Landau sublevels cross the Fermi energy, using a simple model for the spatial dependence of the sublevel energy described by
		\begin{equation}
		E_\pm(x)=\frac{\mu_\mathrm{T}-\mu_\mathrm{B}}{2} g(x) + \frac{\mu_\mathrm{T}+\mu_\mathrm{B}}{2}\pm \frac{U_\mathrm{ex}(x)}{2}
		\end{equation}
		where $\mu_{\mathrm{B(T)}}$ is the chemical potential to the left (right) of the junction, $U_\mathrm{ex}(x)$ is the exchange splitting, and $g(x)$ is a function that smoothly changes from -1 to 1 across the PN junction over a distance $W$ that is similar to the distance of the graphene to the gates. For simplicity, from now on we neglect the spatial dependence of $U_{\mathrm{ex}}$. To link the chemical potentials to filling factors, we note that in general $\mu = \int_{0}^{\nu} \frac{\delta \mu}{\delta \nu'} d \nu'$, which increases stepwise as Landau levels are filled. For simplicity, we make the approximation $\mu=\frac{E_\mathrm{C}}{2}\mathrm{sgn}(\nu)\sqrt{|\nu|}$, where $E_\mathrm{C} = \sqrt{2e\hbar B v_\mathrm{F}^2}$ and $v_\mathrm{F}$ is the Fermi velocity \cite{Zhang2005}. Defining $f_\pm(x)=\frac{4E_\pm(x)}{E_C}$, and limiting ourselves to the case of $\nu_\mathrm{T} < 0$ and $\nu_\mathrm{B} >0$ relevant in our PN measurements, we get 
		\smallskip
		\begin{equation}
		f_\pm(x) = -(\sqrt{-\nu_\mathrm{T}}+\sqrt{\nu_\mathrm{B}})g(x) + \sqrt{\nu_\mathrm{B}}-\sqrt{-\nu_\mathrm{T}} \pm u_\mathrm{ex}
		\end{equation} \smallskip
		where we have defined $u_\mathrm{ex} = \frac{4U_\mathrm{ex}}{E_\mathrm{C}}$.
		The location of an edge channel is then obtained by solving $f(x) = 0$, so that the distance between two edge channels is given by
		\begin{equation}
		\Delta_x = g^{-1} \bigg[ \frac{\sqrt{\nu_\mathrm{B}}-\sqrt{-\nu_\mathrm{T}}+u_\mathrm{ex}}{\sqrt{-\nu_\mathrm{T}}+\sqrt{\nu_\mathrm{B}}} \bigg] - g^{-1}\bigg[\frac{\sqrt{\nu_\mathrm{B}}-\sqrt{-\nu_\mathrm{T}}-u_\mathrm{ex}}{\sqrt{-\nu_\mathrm{T}}+\sqrt{\nu_\mathrm{B}}} \bigg]
		\label{eq:1.3}
		\end{equation}\smallskip
		For $g(x)$ we use a logistic function of width $W$:
		\begin{equation}
		g(x)=\frac{2}{1+e^{-\frac{X}{W}}}-1.
		\end{equation}
		This has the inverse
		\begin{equation}
		g^{-1}(y)=W\mathrm{ln}\bigg[\frac{1+y}{1-y}\bigg].
		\label{eq:1.5}
		\end{equation}
		Combining Eqs.~\ref{eq:1.3} and \ref{eq:1.5}, we find the distance between the two edge channels
		\begin{equation}
		\Delta_x = W \mathrm{ln} \bigg[ \frac{\sqrt{\nu_\mathrm{B}}+u_\mathrm{ex}/2}{\sqrt{\nu_\mathrm{B}}-u_\mathrm{ex}/2}
		\frac{\sqrt{-\nu_\mathrm{T}}+u_\mathrm{ex}/2}{\sqrt{-\nu_\mathrm{T}}-u_\mathrm{ex}/2} \bigg]
		\end{equation}
		To calculate the plot of Fig.~2E, we now assume a MZ interferometer with 50/50 beam splitters and calculate the conductance using
		\begin{equation}
		g = 0.5+0.5 \mathrm{cos} \bigg( 2\pi \frac{B L \Delta_x}{\Phi_0} \bigg)
		\end{equation}
		In Fig.~2E, we used \emph{B} = 9 T, \emph{L} = 1.2 $\mu$m, $\Phi_0 = h/e$, W = 52 nm and, to qualitatively resemble the data, $u_{\mathrm{ex}}$=0.4.
		
		
		\subsection{Supplementary Note 2. Random scattering model for a MZ interferometer at a graphene PN junction}
		
		In this note we provide a background discussion of the electronic wavefunctions corresponding to the edge channels that form an interferometer along the PN junction. We then analyze the probability to find a particular visibility of the Mach-Zehnder conductance oscillations for an interferometer that has beamsplitters described by random transmission/reflection matrices. The resulting cumulative probability distribution function for the visibility is shown as the theoretical curve in Fig.~2G of the main text.\\
		
		\noindent \underline{Straight junction of infinite length}\\
		Let us first consider the case of an infinite, translationally-invariant PN junction aligned with the \emph{y}-axis. The overall problem can be formulated in terms of a solution to the time-independent single-particle Schr{\"o}dinger equation at an energy \emph{E} equal to the Fermi energy.  We may write this solution in the form $\psi (x,y,\tau)$, where $\tau$ is a valley index, and we consider only one spin state. We work in a gauge where the vector potential is parallel to the junction. Then, if there are just two edge states at the junction, we may write, in the vicinity of the junction
		\begin{equation}
		\psi (x,y,\tau) = c_1 \mathrm{e}^{ik_1y} \varPhi_1(x,\tau) + c_2 \mathrm{e}^{ik_2y} \varPhi_1(x,\tau)
		\end{equation}
		where $k_j$, for \emph{j} = 1,2, are the two eigenvalues of the translation operator, $\varPhi_j$ are the corresponding eigenvectors, and $c_j$ are arbitrary constants. It will be convenient to choose the normalizations of $\varPhi_j$ so that
		\begin{equation}
		u_j \int_{-\infty}^{\infty} dx \sum_{\tau} |\varPhi_j|^2 = 1,
		\end{equation}
		where $u_j = (\mathrm{d}k_j/\mathrm{d}E)^{-1}$ is the velocity of the edge mode.
		
		In the simplest model that we have in mind, the functions $\varPhi_j$ will have the approximate form
		\begin{equation}
		\varPhi_j (x,\tau) \approx \chi_j (\tau) \mathrm{exp} [-(x-x_j)^2/(2l_\mathrm{B}^2)],
		\end{equation}
		where $l_B$ is the magnetic length, $\chi_j$ is the center of gravity of the state j, and the spinor $\chi_j$ is a function of the valley index $\tau$. The separation $\Delta_x = x_2 - x_1$ may be written as
		\begin{equation}
		\Delta_x = U_{\mathrm{ex}}/V',
		\end{equation}
		where $V'$ is the gradient of the electrostatic potential and $U_\mathrm{ex}$ is the exchange splitting, which we assume to be constants in the vicinity of the p-n junction. The separation $\Delta_x$ is related to the momentum difference $k_2-k_1$ by
		\begin{equation}
		\Delta_x = |k_2-k_1|l^2_{\mathrm{B}}
		\end{equation}
		
		The exchange splitting in Eq. (2.4) results from a term in the Hartree-Fock Hamiltonian of form:
		\begin{equation}
		H_{\mathrm{ex}} = \frac{U_\mathrm{ex}}{2} \hat{n} \cdot \vec{\tau},
		\end{equation}
		where $\hat{n}$ is a three-component unit vector and $\vec{\tau}$ are the three Pauli matrices, acting on the valley index $\tau$. In an approximation where one neglects valley-dependent electron-electron interactions, which only occur when two electrons are very close together, there is nothing to pick out one particular orientation of $\hat{n}$ over another. Nevertheless, the exchange splitting may be greater than zero. The choice of $\hat{n}$ at a particular PN junction will then depend on small symmetry breaking terms, which we will not attempt to predict, and it could vary as one moves along the junction. The spinors $\varPhi_j$ are the eigenstates of $\hat{n} \cdot \vec{\tau}$.
		
		The magnitude of the exchange splitting must be determined self-consistently, based on the local difference in the occupation of the two valley states. On the microscopic level, we would expect that the exchange potential is not actually a constant over the range of the width of the PN junction, so the quantity $U_\mathrm{ex}$ in the above equations should be taken as an average over a region covered by the wave functions $\varPhi_j$.
		
		It is straightforward to generalize the discussion of a translationally invariant junction to a situation where properties of the junction, including its orientation, may vary adiabatically along its length. (“Adiabatically” means that the distance scale for changes along the length of the junction should be large compared to $(k_2-k_1)^{-1}$.) We choose a gauge where the vector potential is always oriented parallel to the junction at the position of the junction. We may now write
		\begin{equation}
		\psi (x,y,\tau) = c_1 \mathrm{e}^{i \varphi_1 (y)} \varPhi_1(x,\tau) + c_2 \mathrm{e}^{i \varphi_2 y} \varPhi_2(x,\tau)
		\end{equation}
		\begin{equation}
		\psi_j(y) = \int_0^y k_j(y')\mathrm{d}y'
		\end{equation}
		where \emph{y} is the distance along the edge and $x$ is measured in the local perpendicular direction. The fact that the magnitudes $|c_j|$ are independent of $y$ is a consequence of conservation of current and our choice for the normalization of $\varPhi_j$.\\
		
		\noindent \underline{Junction connected to sample edges}\\
		Now we consider a PN junction of length $L$, connected to the sample boundaries at its two ends as depicted in Fig.~1A. We assume that there is a single chiral edge state at any segment of the boundary, and we assume that the edge states at $y = 0$ flow into the junction while the edge states at $y = L$ flow away from the junction.  The portions of the wave function $\psi$ incident at $y = 0$ may be characterized by complex amplitudes $a_1$ and $a_2$, such that $|a_1|^2$ and $|a_2|^2$ are, respectively, the currents incident from the left and from the right. Similarly, we may characterize the outgoing wave function by amplitudes $b_1$ and $b_2$ for electrons moving to the right and left, respectively, away from the end of the junction at $y = L$.
		The amplitudes $c_j$ for the wave function along the PN junction will be related to the amplitudes $a_j$ by a 2 x 2 unitary matrix $S^{(0)}$, whose form will depend on details of the sample in the region where the junction meets the edge.  Similarly, we may define a matrix $S^{(L)}$, which relates the outgoing amplitudes $b_j$ to the amplitudes $c_j e^{i \varphi_j (L)}$ at the end of the PN junction. The outgoing amplitudes $b_j$ will then be related to the incoming amplitudes $a_j$ by a matrix $N$, which we may write as
		\begin{equation}
		N = S^{(L)} M S^{(0)},
		\end{equation}
		where $M$ is a diagonal matrix with elements $M_{jj'} = \delta_{jj'} \mathrm{e}^{i \varphi_j (L)} $. Suppose that there is incident beam impinging on the junction from the left, so that $a_2 = 0$. Defining the transmission coefficient $T$ as the probability for an electron to wind up in the right-moving state after leaving the junction, we see that 
		\begin{equation}
		T = |N_{11}|^2
		\end{equation}
		
		Let $\hat{n}_\alpha$ be the unit vector on the Bloch sphere that corresponds to the two-component unit vector $ a_j \equiv S^{(0)}_{j,1} $, and let $\hat{n}_\beta$ be the unit vector on the Bloch sphere that corresponds to the two-component unit vector $\beta \equiv \Big[ \Big( S^{(L)} \Big)^{-1} \Big] _{j,i} $. Let $R_{\varphi}$ by the $O(3)$ matrix that represents a rotation by an angle $\varphi$ about the z-axis, where
		\begin{equation}
		\varphi = \varphi_2(L) - \varphi_1(L) = \int_0^L (k_2-k_1) \mathrm{d}y
		\end{equation}
		Then we have
		\begin{equation}
		T = \frac{1+\hat{n}_\beta \cdot R_\varphi \cdot \hat{n}_\alpha}{2}
		\end{equation}
		
		If we describe the vectors $\hat{n}_\alpha$ and $\hat{n}_\beta$ by their polar coordinates, $(\theta, \varphi)$, and define $\varphi_0 = \varphi_\beta - \varphi_\alpha$ then we obtain
		\begin{equation}
		T = C + D \mathrm{cos} (\varphi + \varphi_0)
		\label{eq:2p13}
		\end{equation}
		\begin{equation}
		C = \frac{1 + \mathrm{cos} \theta_\alpha \mathrm{cos} \theta_\beta}{2}, \mathrm{and}\
		D = \frac{\mathrm{sin} \theta_\alpha \mathrm{sin} \theta_\beta}{2}
		\end{equation}
		where we recognize Eq. 1 of the main text with $|t_1|^2 = \cos^2\frac{\theta_\alpha}{2}$ and $|r_2|^2 = \cos^2\frac{\theta_\beta}{2}$. If the length $L$ is large, the phase $\varphi$ will change by a large amount when we vary the magnetic field by an amount that is still too small to affect the other parameters in the above equation. Thus $T$ will oscillate between maximum and minimum values given by
		\begin{equation}
		T_\mathrm{max} = C+D \ \mathrm{and} \ T_\mathrm{min} = C-D
		\end{equation}
		If we define the visibility by
		\begin{equation}
		V = \frac{T_\mathrm{max}-T_\mathrm{min}}{T_\mathrm{max}+T_\mathrm{min}}
		\end{equation}
		then we find
		\begin{equation}
		V = \frac{D}{C} = \frac{\mathrm{sin}\theta_\alpha \mathrm{sin}\theta_\beta}{1+\mathrm{cos}\theta_\alpha \mathrm{cos}\theta_\beta}
		\end{equation}
		
		\noindent \underline{Random scattering model}
		
		Since the matrices $S^{(0)}$ and $S^{(L)}$ depend on details that we do not know how to calculate, we will consider a model in which the matrices are random matrices in the group $U(2)$. In this case, the unit vectors $\hat{n}_\alpha$ and  $\hat{n}_\beta$ will be randomly distributed on the Bloch sphere.
		We wish to calculate the probability $P(\epsilon)$ that the visibility $V$ is greater than 1 - $\epsilon$. It is useful to change variables to 
		\begin{equation}
		x = \frac{\mathrm{cos}\theta_\alpha+ \mathrm{cos}\theta_\beta}{2},
		y = \frac{\mathrm{cos}\theta_\alpha- \mathrm{cos}\theta_\beta}{2}
		\end{equation}
		The probability $P(\epsilon)$ is then given by:
		\begin{equation}
		P(\epsilon) = \frac{A(\epsilon)}{2}
		\end{equation}
		where $A$ is the area of the $x-y$ plane that satisfies the constraints
		\begin{equation}
		-1 < x < 1, -1 < y+x < 1, -1 < y-x < 1,
		\label{eq:constraints}
		\end{equation}
		
		\begin{equation}
		\begin{split}
		&\epsilon > \frac{C-D}{C} = \\ &\frac{1+x^2-y^2-[(1-y^2-x^2-2xy)(1-y^2-x^2+2xy)]^{(1/2)}}{1+x^2-y^2}
		\end{split}
		\label{eq:expand}
		\end{equation}
		
		Ideally, we should compute $A(\epsilon)$ numerically, which we do to calculate the theoretical curve displayed in Fig.~2G. However, we can make an analytic approximation, which should be valid in the limit of small $\epsilon$. In this limit, we can expand the right hand side of Eq.~\ref{eq:expand} and replace it by the constraint
		\begin{equation}
		\epsilon > \frac{2x^2}{(1-y^2)^2} 
		\end{equation}
		Since this constraint forces $|x|< (\frac{\epsilon}{2})^{1/2}$, when $\epsilon$ is small we can replace the constraints (Eqn.~\ref{eq:constraints}) by
		\begin{equation}
		-1 < y < 1 
		\end{equation}
		The integral is now simple to carry out, giving us
		\begin{equation}
		P(\epsilon) \approx \int^1_{-1} \mathrm{d}y (1-y^2)(\epsilon/2)^{1/2} = \bigg(\frac{8\epsilon}{9} \bigg) ^{1/2} 
		\end{equation}
		
		
		\subsection{Supplementary Note 3. Conductance of two Mach-Zehnder interferometers in series}
		
		In this note, we analyze the conductance values observed in the different regions indicated in Fig.~4C of the main text. As discussed in the main text, we attribute the conductance observed in regions II and III to the presence of one interferometer at each of the PN junctions, and the conductance in region IV to the presence of two interferometers at each of the PN junctions. We thus analyze the expected conductance of MZ interferometers that are connected in series, as depicted in Supplementary Figure~4A.
		
		We first assume that we have only one interferometer at each of the PN junctions (corresponding to regions II and III in Fig.~4C). We also assume that phase coherence is lost in the region between the two interferometers, as the edge channels run along several microns of vacuum edge. The transmission through the two interferometers is given by
		\begin{equation}
		\begin{split}
		T = T_1 T_2 \sum ^\infty_{k=0} [(1-T_2)(1-T_1)]^k & = T_1T_2 \frac{1}{1-(1-T_2)(1-T_1)} \\
		& = \frac{1}{\frac{1}{T_1} + \frac{1}{T_2}-1}
		\end{split}
		\label{eq:3p1}
		\end{equation}
		where $T_i$ is the transmission probability through interferometer $i$. Recalling Supplementary Eq.~\ref{eq:2p13}, the transmission through a MZ interferometer is given by
		\begin{equation}
		T_i = C_i +D_i \mathrm{cos} \Big( \varphi^{(i)} + \varphi_0^{(i)} \Big) 
		\end{equation}
		Likewise, the transmission through an NPN device with two interferometers (that are independent because of their opposite spin) at each PN interface is given by
		\begin{equation}
		\begin{split}
		T& = T^{\uparrow} +T{\downarrow} \\
		& = \frac{1}{\frac{1}{T_1^{\uparrow}} + \frac{1}{T_2^\uparrow}-1}
		+ \frac{1}{\frac{1}{T_1^{\downarrow}} + \frac{1}{T_2^\downarrow}-1}
		\end{split}
		\label{eq:3p3}
		\end{equation}
		To compare the average conductance observed in regions II, III, and IV of Fig.~4C of the main text to expectations based on this model, we now assume 50/50 beam splitters and average Eqs.~\ref{eq:3p1} and \ref{eq:3p3} over $\varphi^{(i=1,2)}$. The results are shown in Supplementary Figure 4C. We see a reasonable agreement. However, we note that it is unclear why one should expect 50/50 beamsplitters. 
		
		
		\subsection{Supplementary Note 4. Analyzing the gate-length dependence of the Mach-Zehnder oscillation frequencies observed in the NPN measurements on device 2.}
		
		To determine the location of the beamsplitters of our Mach-Zehnder interferometers, we analyze the frequency of the conductance oscillations in the top gate/back gate sweeps of our NPN device (device 2) for each of the five top gates. As discussed in the main text, we analyze the frequency of the oscillations observed in the NPN conductance data such as those shown in Fig.~4C by focusing on the region of large filling factors ($|\nu_{\mathrm{B,T}}| \gg 1 $). In this regime, we can linearize the function $g(x)$ in Eq.~\ref{eq:1.3}, so that it follows that for a given value of $(\sqrt{\nu_\mathrm{B}}-\sqrt{-\nu_\mathrm{T}})$, the edge-channel separation varies approximately as $\Delta_x \propto \frac{1}{\sqrt{\nu_\mathrm{B}}+\sqrt{-\nu_\mathrm{T}}}$.
		
		As such, the frequency of the Mach-Zehnder oscillations should be constant as a function of $\frac{1}{\sqrt{\nu_\mathrm{B}}+\sqrt{-\nu_\mathrm{T}}}$. Therefore, for each top gate, we take the NPN conductance data (such as the data shown in Fig.~4C) and plot it against $x = \frac{-1}{\sqrt{\nu_\mathrm{B}}+\sqrt{-\nu_\mathrm{T}}}$ and  $y = (\sqrt{\nu_\mathrm{B}}+\sqrt{-\nu_\mathrm{T}})$ (Supplementary Figure 5). We then divide the measurement range as indicated by the boxes, focusing on the limit $|\nu_{\mathrm{B,T}}| \gg 1$ (i.e, focusing on the right-hand side of the plots), and calculate the absolute value of the Fourier transform with respect to the $x$-coordinate for all data traces within each box. For each box we average these Fourier transforms over the $y$-coordinate.  In order to convert from frequency to gate length, we multiply the $x$-coordinates in each box by a scale factor, which depends on the $y$-coordinate but is the same across all gates for all boxes with the same $y$-coordinate.  The scale factors are chosen so that for TG5, the peak of the Fourier transform occurs at the gate length 22 $\mu$m for each value of $y$. Finally, we average the resulting frequency-axis-normalized spectra (1 for each box) over all boxes. The results are plotted in Fig.~4D of the main text.
		
		
		\subsection{Supplementary Note 5. Gate-defined equilibration studies.}
		
		In this note, we describe how we derive the expected equilibration curves shown in Fig.~5E of the main text (red lines), which are based on the assumption that edge-channels only equilibrate if they have the same spin and run along the physical graphene edge.
		In the measurements of Fig.~5E we apply a voltage $V_\mathrm{IN}$ to the left lead, ground the top lead ($V_\mathrm{G}=0$), and measure the voltage $V_\mathrm{OUT}$ at the right lead. Our goal is to calculate $V_\mathrm{OUT}$ as a function of the side-gate filling factor $\nu_\mathrm{S}$. We keep the filling factor under the left and right top gate equal, calling it $\nu_\mathrm{T}$. Furthermore, we work in the regime where $\nu_\mathrm{B} > \nu_\mathrm{T} \geq 1$, where $\nu_\mathrm{B}$ is the filling factor in the non-top-gated regions, so that we expect $V_\mathrm{OUT}=V_\mathrm{IN}$ if there is no edge-channel equilibration in the central region and under the assumption that the resistances of the leads are zero. 
		
		To calculate the expected $V_\mathrm{OUT}$, we start by assuming an infinite input impedance of our voltmeter, allowing us to relate $V_\mathrm{OUT}$ to the chemical potentials of the edge channels arriving at the right lead as
		\begin{equation}
		V_\mathrm{OUT} = -\frac{\nu_\mathrm{T}^\uparrow \mu_\mathrm{OUT}^\uparrow+\nu_\mathrm{T}^\downarrow \mu_\mathrm{OUT}^\downarrow}{\nu_\mathrm{T}^{\uparrow}+\nu_\mathrm{T}^\downarrow}
		\label{eqn:5p1}
		\end{equation}
		where $\nu_\mathrm{T}^\uparrow$ and $\nu_\mathrm{T}^\downarrow$ are the number of spin-up and spin-down edge channels under the left and the right top gate, with 〖$\nu_T = \nu_\mathrm{T}^\uparrow + \nu_\mathrm{T}^\downarrow$, and $\mu_\mathrm{OUT}^\uparrow$ and $\mu_\mathrm{OUT}^\downarrow$ are the chemical potentials of these channels when they arrive at the right lead. 
		
		We now assume that edge channels only equilibrate if they have the same spin and run along the physical graphene edge. The number of spin-up and spin-down edge channels running along the lower physical edge in the central region is given by $\nu_\mathrm{S}^\uparrow$ and $\nu_\mathrm{S}^\downarrow$, respectively, with $\nu_\mathrm{S} = \nu_\mathrm{S}^\uparrow + \nu_\mathrm{S}^\downarrow$. We thus expect
		\begin{equation}
		\begin{split}
		\mu_\mathrm{OUT}^{\uparrow,\downarrow} & = \frac{\nu_\mathrm{T}^{\uparrow,\downarrow} \mu_\mathrm{IN}+ (\nu_\mathrm{S}^{\uparrow,\downarrow}-\nu_\mathrm{T}^{\uparrow,\downarrow}) \mu_\mathrm{G}} {\nu_\mathrm{S}^{\uparrow,\downarrow}} \\
		\mathrm{for} \ & \Big( \nu_\mathrm{S}^{\uparrow,\downarrow} > \nu_\mathrm{T}^{\uparrow,\downarrow} \Big) \wedge \Big(\nu_{\mathrm{S}} \leq \nu_\mathrm{B} \Big)
		\end{split}
		\label{eqn:5p2}
		\end{equation}
		and 
		\begin{equation}
		\begin{split}
		\mu_\mathrm{OUT}^{\uparrow,\downarrow} & = \frac{\nu_\mathrm{T}^{\uparrow,\downarrow} \mu_\mathrm{IN}+ (\nu_\mathrm{B}^{\uparrow,\downarrow}-\nu_\mathrm{T}^{\uparrow,\downarrow}) \mu_\mathrm{G}}{\nu_\mathrm{B}^{\uparrow,\downarrow}} \\
		\mathrm{for} \ & \Big( \nu_\mathrm{S}^{\uparrow,\downarrow} > \nu_\mathrm{T}^{\uparrow,\downarrow} \Big) \wedge \Big(\nu_\mathrm{S} > \nu_\mathrm{B} \Big)
		\end{split}
		\label{eqn:5p3}
		\end{equation}
		and 
		\begin{equation}
		\mu_\mathrm{OUT}^{\uparrow,\downarrow} = \mu_\mathrm{IN} \ \mathrm{for} \ \Big( \nu_\mathrm{S}^{\uparrow,\downarrow} \leq \nu_\mathrm{T}^{\uparrow,\downarrow} \Big) 
		\label{eqn:5p4}
		\end{equation}
		where $\mu_\mathrm{G}$ and $\mu_\mathrm{IN}$ are the chemical potentials of the edge channels emerging from the top and left lead respectively. Equations \ref{eqn:5p2} to \ref{eqn:5p4}, substituted into Eq.~\ref{eqn:5p1}, describe the expected equilibration curves shown in Fig.~5E of the main text (red lines). However, to get those curves we need to include the effect of the non-zero resistances of the left and top lead, $R_\mathrm{IN}$ and $R_\mathrm{G}$ respectively. 
		The lead resistances are non-zero because of RC filters, wires, and contact resistance. From current conservation at the left and top lead, we get 
		\begin{equation}
		\mu_\mathrm{IN} = -V_\mathrm{IN} \frac{\nu_\mathrm{T} R_\mathrm{G} +R_Q}{\nu_T(R_\mathrm{IN}+R_\mathrm{G})+R_Q}
		\label{eqn:5p5}
		\end{equation}
		and
		\begin{equation}
		\mu_\mathrm{G} = -V_\mathrm{IN} \frac{\nu_\mathrm{T} R_\mathrm{G}}{\nu_\mathrm{T}(R_\mathrm{IN}+R_\mathrm{G})+R_Q}.
		\label{eqn:5p6}
		\end{equation}
		Experimentally, we determine the resistances of the leads by setting $\nu_\mathrm{B} = \nu_\mathrm{T} = 2$ and measuring the quantum Hall resistance plateau using the left and the top lead. The value of this plateau is given by $R=R_Q/2+R_\mathrm{IN}+R_\mathrm{G}$, with $R_Q=h/e^2$, allowing us to extract $R_\mathrm{IN}+R_\mathrm{G}$. We then assume $R_\mathrm{IN}=R_\mathrm{G}$ yielding $R_\mathrm{IN}=R_\mathrm{G} = 4.4 k\Omega$. Using Eqs.~\ref{eqn:5p1} to \ref{eqn:5p6}, we obtain the traces plotted in Fig.~5E of the main text. 
		
		
		\subsection{Supplementary Note 6. Calculating charge densities and filling factors from gate voltages.} 
		
		In this note we describe how we obtain the filling factor values displayed on the axes of our plots. First, we estimate the charge density $n_\mathrm{B}$ in the non-top-gated region using a simple parallel-plate capacitor model via the equation $n_\mathrm{B}=\epsilon_0 \epsilon_\mathrm{B} V_\mathrm{B}/(d_\mathrm{B} e)$, where $\epsilon_0$ is the vacuum permittivity, $\epsilon_\mathrm{B}$ is the dielectric constant of the back-gate dielectric (either hBN or SiO$_2$), $V_\mathrm{B}$ is the applied back-gate voltage, $d_\mathrm{B}$ is the thickness of the back-gate dielectric, and $e$ is the electron charge. To calculate the charge density $n_\mathrm{T}$ in a top-gated region, we use the equation $n_\mathrm{T}=n_\mathrm{B}+\epsilon_0 \epsilon_\mathrm{T} V_\mathrm{T}/(d_\mathrm{T} e)$, where $\epsilon_\mathrm{T}$ is the dielectric constant of the hBN top-gate dielectric, $V_\mathrm{T}$ is the applied top-gate voltage, and $d_\mathrm{T}$ is the thickness of the top-gate dielectric. 
		
		To determine the precise location of the charge neutrality point, we measure the conductance while sweeping both the top and bottom gate. We locate the center of the $\nu_\mathrm{B}=0$ plateau, and if it is shifted from $V_\mathrm{B}=0$ by $V_{\mathrm{B},\mathrm{off}}$ we take this into account by substituting $V_\mathrm{B} \rightarrow V_\mathrm{B}-V_{\mathrm{B,off}}$ in our equation for $n_\mathrm{B}$. We use a similar procedure for $n_\mathrm{T}$. As described in the main text, we then calculate the filling factor $\nu_{(\mathrm{T,B})}$ using the equation $\nu_{(\mathrm{T,B})} =(h/eB)n_{(\mathrm{T,B})}$  where $B$ is the magnetic field, and $h$ is Planck's constant. 
		
		Due to quantum capacitance effects, these filling factors do not precisely correspond with the positions where we see robust plateaus in the quantum Hall regime. To correct for this, we tune to an NN$'$ configuration (defined by $\nu_\mathrm{T}>0$ and $\nu_\mathrm{B}>0$) where we are able to identify the locations of particular filling factors by the observation of robust quantum Hall plateaus. We line up the filling factor on the axis with the corresponding plateau by multiplying by a constant factor $C_\mathrm{q}$ that accounts for the quantum capacitance. Then, to determine the location of the filling factors in the NP regime, we assume that the quantum Hall plateaus induced on the electron side are of the same size as the quantum Hall plateaus induced on the hole side, and multiply the filling factor obtained from the parallel-plate capacitor model by the same  $C_\mathrm{q}$ factor.  We note that exchange-split levels often span a smaller density range than those separated by a cyclotron energy gap (i.e. the $\nu=1$ plateau is smaller than the $\nu=2$ plateau), introducing a small uncertainty into this procedure.
		
		\begin{figure*}[t]
			\includegraphics[width=0.7\textwidth]{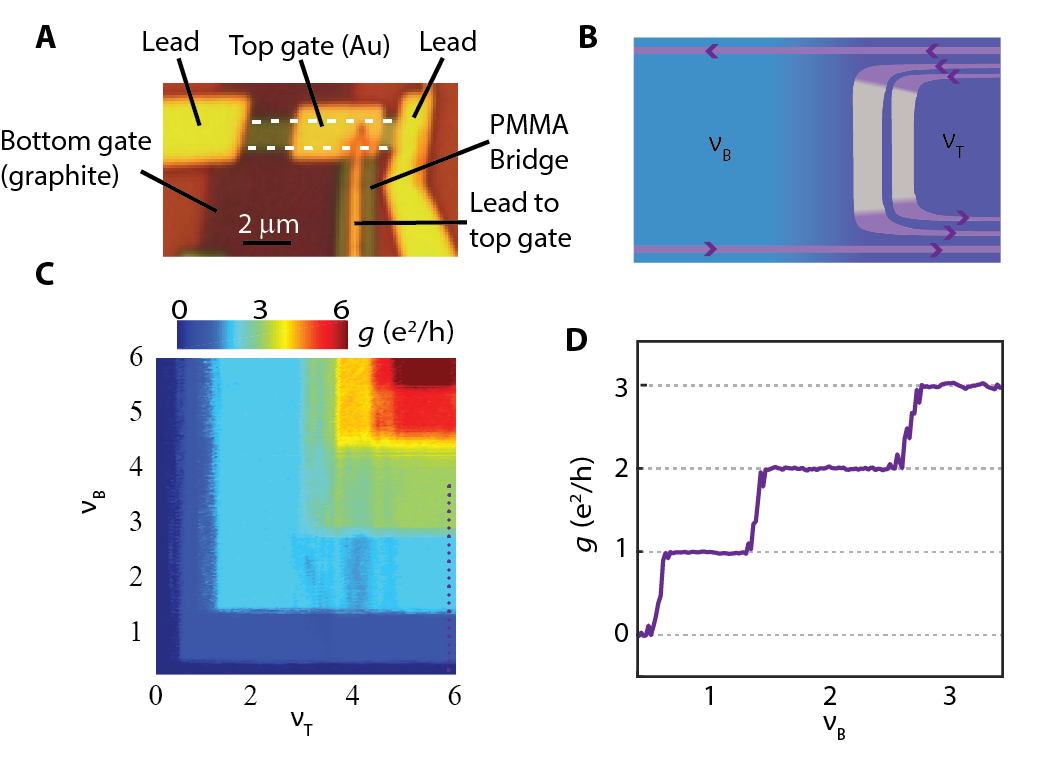}
			\caption{\textbf{Characterization of device 1 in the regime where $\nu_\mathrm{B} > 0$ and $\nu_\mathrm{T} > 0$ (which we call the NN$'$ regime).} \textbf{(A)} Optical microscope image of device 1: a triple-gated, hexagonal boron nitride-encapsulated monolayer of graphene. This image corresponds to the schematic in Fig.~1B of the main text. The encapsulated graphene is outlined by the white dashed lines. We tune the filling factor $\nu_\mathrm{T}$ under the top gate using both the graphite bottom gate and the Au/Cr top gate. We tune the filling factor $\nu_\mathrm{B}$ in the region to the left of the top gate using the bottom gate only. The device sits on a 285 nm $\mathrm{SiO}_2$/Si global back gate that we use to strongly dope the graphene leading up to the right lead, thus reducing the contact resistance. A bridge of hard-baked PMMA supports the lead contacting the top gate and prevents shorting of this lead to the graphene. \textbf{(B)} A schematic illustration of the edge states present in the system in the NN$'$ regime, with ($\nu_\mathrm{B}$, $\nu_\mathrm{T}$) = (1,3) as an example. In the NN$'$ regime, the conductance is given by min($\nu_\mathrm{B}$, $\nu_\mathrm{T}$). \textbf{(C)} Two-terminal conductance in the NN$'$ regime. The Si back gate is set to 60 V. \textbf{(D)} Line trace corresponding to the dotted purple line in (B). The observation of conductance quantization in steps of $e^2/h$ confirms that the spin and valley degeneracy is fully lifted in the zLL.}
			\label{fig:Supp1}
		\end{figure*}
		
		\begin{figure*}[t]
			\includegraphics[width=1.0\textwidth]{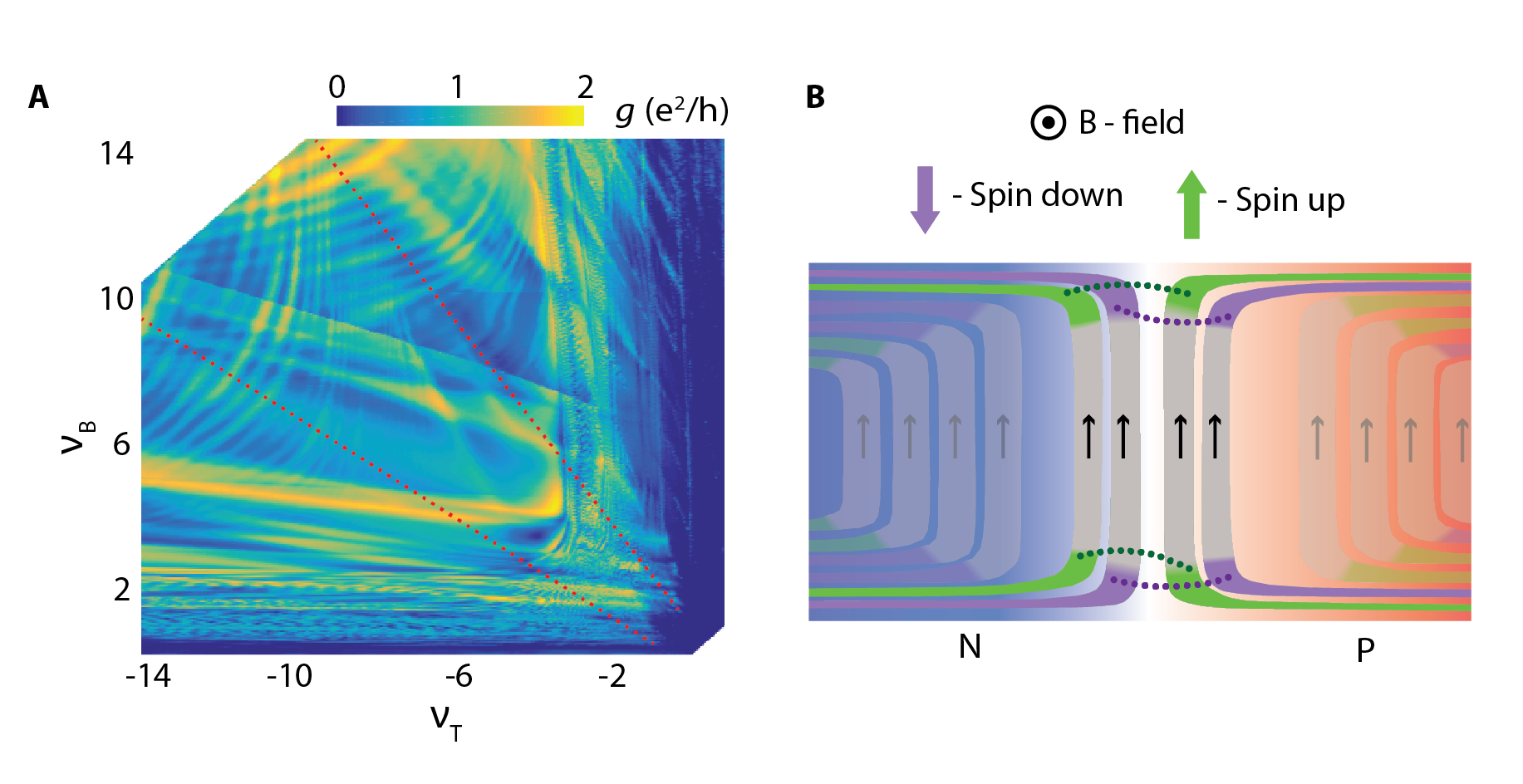}
			\caption{\textbf{Two-terminal conductance of device 1 in the PN regime (in which $\nu_\mathrm{B} > 0$ and $\nu_\mathrm{T} < 0$) at \emph{B} = 4 T and large filling factors.} \textbf{(A)} We observe two hyperbola-shaped sets of conductance oscillations (the red dotted lines guide the eye through the centers of these hyperbolas). As discussed in the main text, our data indicates that these Mach-Zehnders are formed by the two pairs of same-spin edge channels belonging to the zLL. We observe that the conductance oscillates approximately between 0 and $2e^2/h$, even at large filling factors, indicating that two Mach-Zehnder interferometers mediate transport across the PN junction even when there are many edge channels in the system. \textbf{(B)}  Schematic that depicts edge channels belonging to higher LLs that do not communicate across the junction or with the zLL, presumably because of their larger spatial separation. }
			\label{fig:Supp2}
		\end{figure*}
		
		\begin{figure*}[t]
			\includegraphics[width=0.5\textwidth]{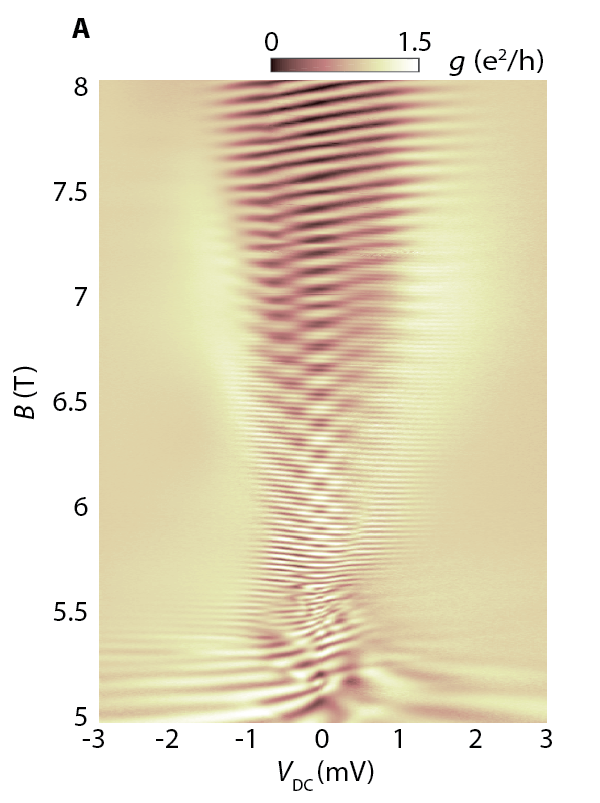}
			\caption{\textbf{The effect of a DC bias on the differential conductance of a PN junction.} \textbf{(A)} Two-terminal conductance of device 1 as a function of magnetic field \emph{B} and DC voltage bias $V_\mathrm{DC}$. At \emph{B} = 8 T, we have $\nu_\mathrm{B}$ = 1 and $\nu_\mathrm{T}$ = −2. Different spatial derivatives of the energies of the two Landau sublevels forming an interferometer (see e.g., Fig.~2D of the main text) can lead to an energy-dependent inter-edge-channel distance $\Delta_x$, which results in a differential conductance that depends on $V_\mathrm{DC}$: if $V_\mathrm{DC}$ is applied asymmetrically, as in our measurements (with the chemical potential of the left channel raised to $V_\mathrm{DC}$ and that of the right channel remaining at 0V), the differential conductance is given by $ g \sim \cos⁡[\frac{2\pi BL}{\Phi_0}  \Delta_x (V_\mathrm{DC})]$. In this case it is clear that a change in $\Delta_x$ caused by a change in $V_\mathrm{DC}$ can be compensated for by a change in \emph{B}, consistent with the diagonal stripes of constant differential conductance observed in the region around \emph{B} = 8 T and in Fig.~3A of the main text. If the bias is somehow symmetrized, due to e.g. electron-electron interactions \cite{McClure2009}, the chemical potential of the left (right) channel equals $\frac{V_\mathrm{DC}}{2}$ ($\frac{-V_\mathrm{DC}}{2}$), and correspondingly $g \sim \cos[\frac{2\pi BL}{\Phi_0}  \Delta_x (\frac{V_\mathrm{DC}}{2})]$+ $\cos[ \frac{2\pi BL}{\Phi_0}  \Delta_x (\frac{-V_\mathrm{DC}}{2})]$. This may lead to more complex behavior such as the checkerboard patterns observed around \emph{B} = 6 T \cite{McClure2009}. We note that a bias-dependent electrostatic gating effect may also change the inter-channel distance \cite{Bieri2009} and correspondingly lead to a bias-dependent differential conductance.}
			\label{fig:Supp3}
		\end{figure*}
		
		\begin{figure*}[t]
			\includegraphics[width=0.75\textwidth]{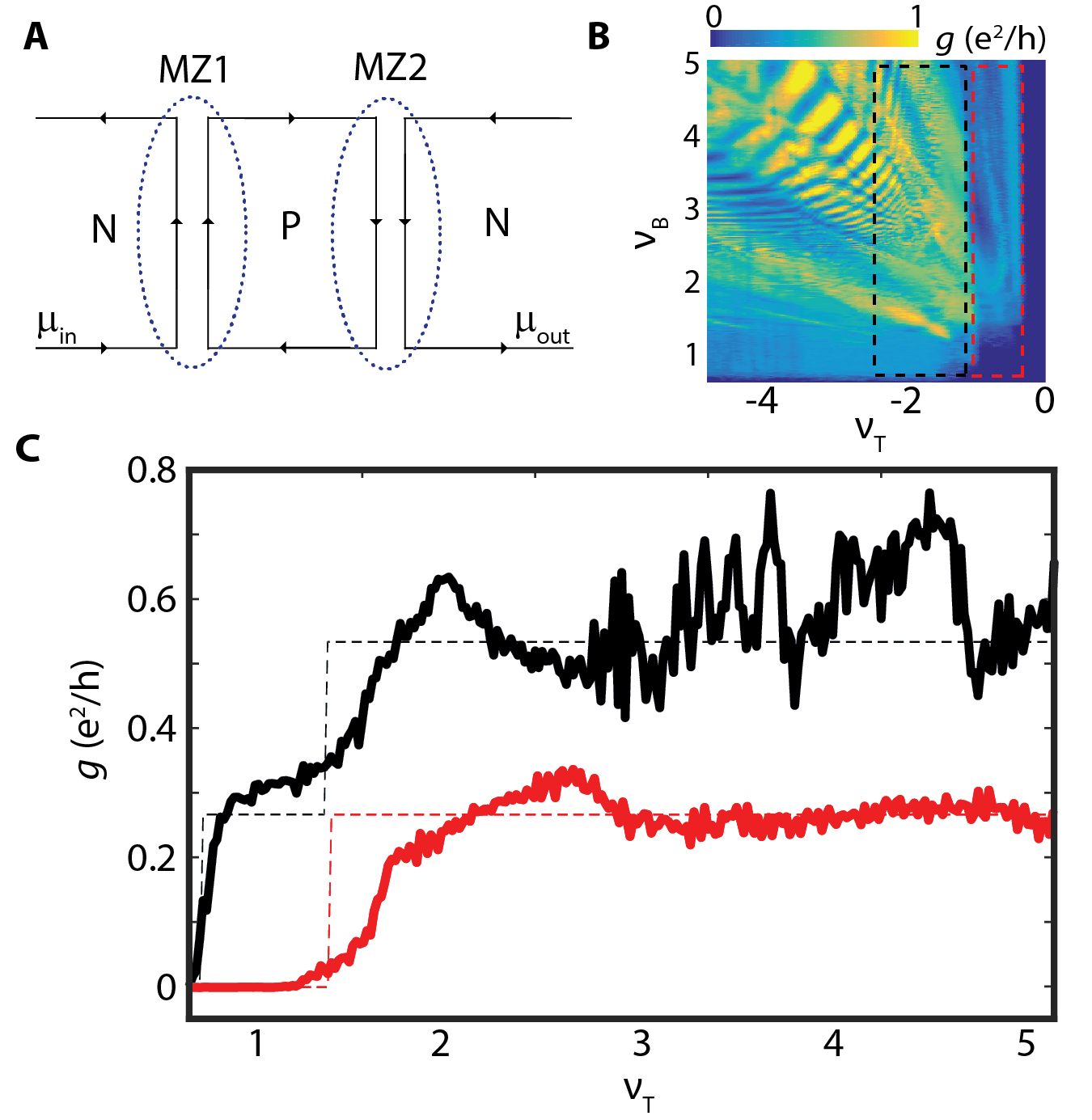}
			\caption{\textbf{Analyzing the average conductance observed in NPN measurements on device 2.} \textbf{(A)}  Schematic of two PN junctions in series (MZ1 and MZ2), as formed in our NPN device. $\mu_\mathrm{in}$ is the chemical potential of the edge entering the first interferometer and $\mu_\mathrm{out}$  is the chemical potential exiting the second interferometer. \textbf{(B)} Two-terminal conductance measurement as a function of back-gate and top-gate filling factors $\nu_\mathrm{B}$ and $\nu_\mathrm{T}$, measured across top gate 1. The red (black) dashed box indicates a region with one (two) edge channel(s) in the top-gated region. \textbf{(C)}  The red (black) data corresponds to the measured conductance within the red (black) dashed box in (B), averaged over $\nu_\mathrm{T}$. The dashed lines indicate the expected average conductance corresponding to 0, 1, or 2 interferometers formed at each of the PN interfaces, assuming 50/50 beamsplitters as discussed in Supplementary Note 3.}
			\label{fig:Supp1}
		\end{figure*}
		
		\begin{figure*}[t]
			\includegraphics[width=1.0\textwidth]{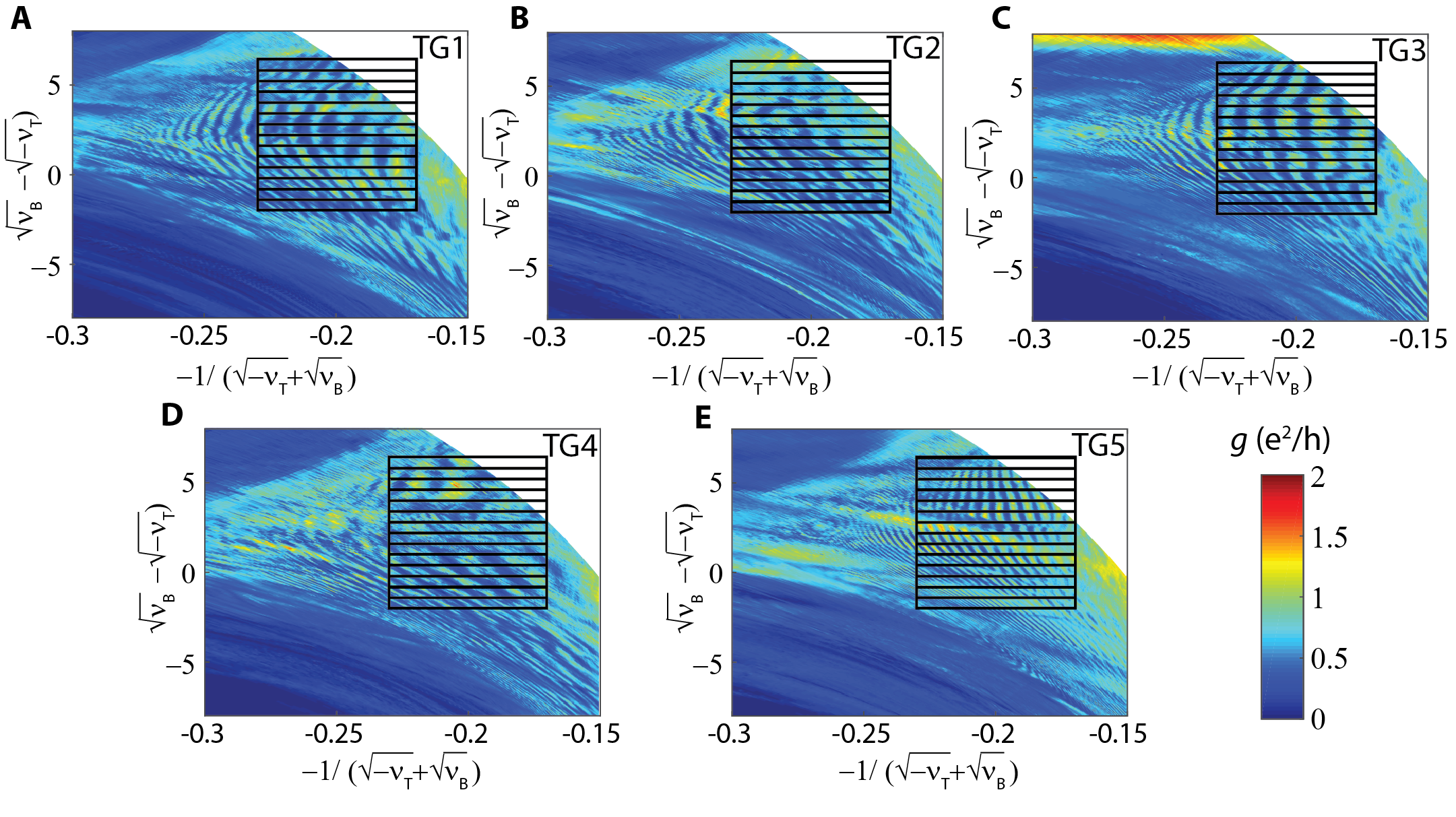}
			\caption{\textbf{Analyzing the gate-length dependence of the Mach-Zehnder oscillation frequencies observed in NPN devices.} \textbf{(A-E)} The conductance maps for all top gates (TG1-TG5) at \emph{B} = 8 T, plotted in a transformed coordinate system. We address each top gate individually by using the appropriate leads. An image of the device is shown in Fig.~4A of the main text. The lengths of the top gates are shown in Fig.~4B of the main text. Note that top gate 5 is the longest and correspondingly shows the fastest conductance oscillations. The boxes indicate the regions in which we take Fourier transforms of the data to compare the frequency of the observed Mach-Zehnder oscillations between the different gates, resulting in Fig.~4D of the main text (see Supplementary Note 4). 	\\}
			\label{fig:Supp1}
		\end{figure*}

		\begin{figure*}[t]
			\includegraphics[width=1.0\textwidth]{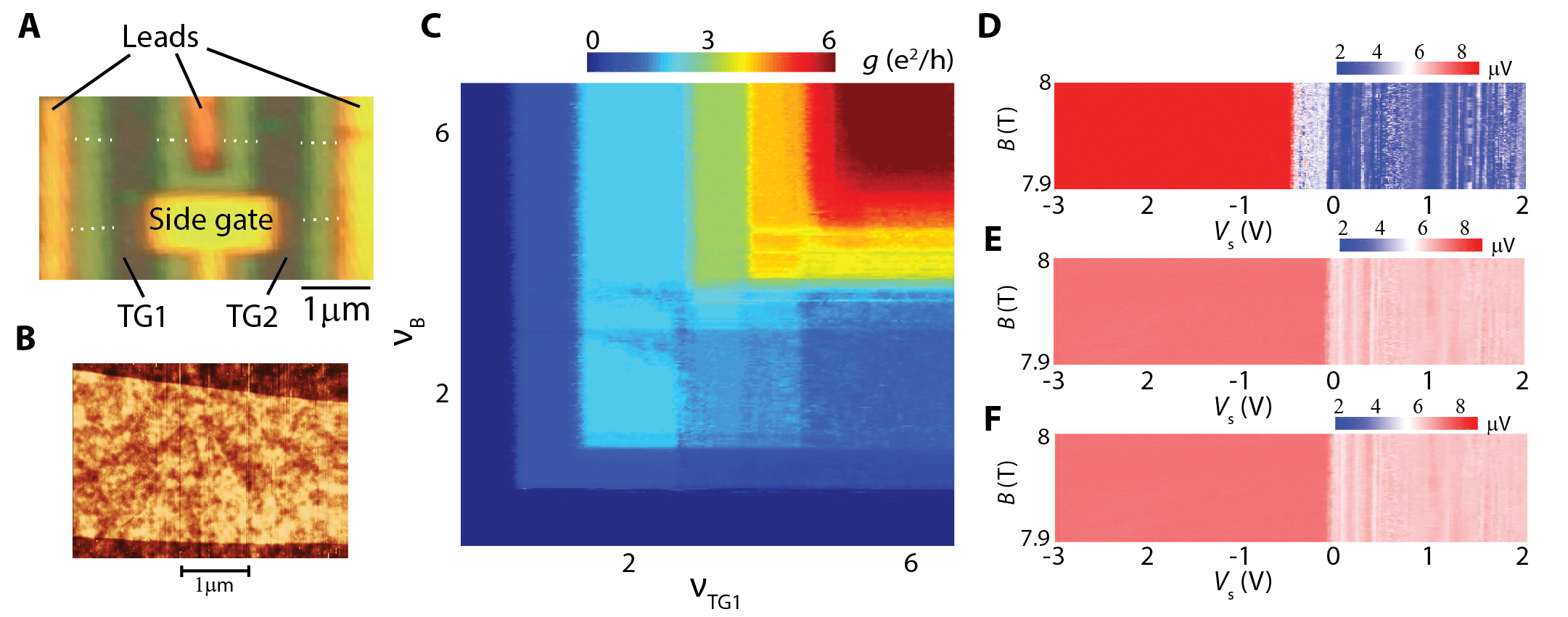}
			\caption{\textbf{Device 3: verifying the presence of broken-symmetry quantum Hall states and measurements of edge channel equilibration as a function of magnetic field.} \textbf{(A)} Optical image of device 3. The dotted line outlines the graphene. \textbf{(B)} AFM image of the clean, hBN-encapsulated graphene flake used for device 3. \textbf{(C)} Two-terminal conductance measured across top gate 1 (TG1) using the left and middle lead, as a function of the filling factor $\nu_\mathrm{TG1}$ under TG1 and the filling factor $\nu_\mathrm{B}$ in the non-top-gated region. Conductance plateaus that are present for all integers from $\nu$ = 1 to  $\nu$ = 6 confirm that the spin and valley degeneracy of the Landau levels is lifted. These plateaus are also present in a similar measurement across TG2 (not shown). \textbf{(D-F)} Equilibration measurements as a function of B and the side-gate voltage $V_\mathrm{S}$, as described in the main text. The two top gates are set at $\nu_\mathrm{T}$=1 for (D) $\nu_\mathrm{T}$=2 for (E) and $\nu_\mathrm{T}$=3 for (F)}
			\label{fig:Supp1}
		\end{figure*}
		
	\end{document}